\documentclass[a4paper,12pt]{article}
\usepackage{amsmath,amssymb,epsfig}
\usepackage{setspace}
\usepackage{graphicx}
\usepackage{caption}
\usepackage{subcaption}

\newcommand{\bra}[1]{\langle #1|}
\newcommand{\ket}[1]{|#1\rangle}

\title{Action with Acceleration II: Euclidean Hamiltonian and Jordan Blocks}
\author{\normalsize
\begin{tabular}[t]{c@{\extracolsep{3em}}c@{\extracolsep{3em}}c@{\extracolsep{3em}}c}
\large Belal E. Baaquie \\
Department of Physics, National University of Singapore \\
2 Science Drive 3, Singapore 117542, Singapore\\
phybeb@nus.edu.sg
\end{tabular}}
\date{\today}
\begin{document}
\maketitle
\begin{abstract}
The Euclidean action with acceleration has been analyzed in \cite{bebeupath}, hereafter cited as reference I, for its Hamiltonian and path integral. In this paper, the state space of the Hamiltonian is analyzed for the case when it is pseudo-Hermitian (equivalent to a Hermitian Hamiltonian), as well as the case when it is inequivalent. The propagator is computed using both  creation/destruction operators as well as the path integral. A state space calculation of the propagator shows the crucial role played by the dual state vectors that yields a result impossible to obtain from a Hermitian Hamiltonian acting on a Hilbert space. When it is not pseudo-Hermitian, the Hamiltonian is shown to be a direct sum of Jordan blocks.
\end{abstract}
\section{Introduction}
The action with acceleration arises in many different problems and provides a new class of quantum mechanical systems; the action in Euclidean time is discussed in paper I. The state space of a single quantum particle described by the acceleration action is function of two distinct degrees of freedom, namely velocity $v$ and position $x$, with the two being related by a constraint equation; the state vector is given by $\psi(x,v)$, quite unlike a quantum particle with an action having only the velocity term (without acceleration) for which the state vector is given by  $\psi(x)$.

Pseudo-Hermitian Hamiltonians have been studied in \cite{mostafazadeh}, \cite{qing}. The analysis of the Euclidean Hamiltonian is a continuation of the ground-breaking analysis of Bender and Mannheim \cite{mannheim1}, \cite{mannheimmain} for Minkowski time; the Euclidean case has some new features that are absent for Minkowski time, the most important being the positivity of the Euclidean state space.

The paper is organized as follows. In Section \ref{shoham} the equivalent simple harmonic oscillator $H_0$ is defined for Euclidean Hamiltonian $H$; in Section \ref{hamacclnact} all the eigenfunctions of $H$ and $H^\dagger$ are obtained; in Section \ref{excitedstham} a few of the excited states $H$ as well as their duals states; in Section \ref{sec:proppseudo} the propagator is derived using the creation and destruction operators and in Section \ref{propstatesp} a state space derivation is given.  The case of the Hamiltonian being inequivalent to a Hermitian Hamiltonian is analyzed in Sections \ref{sec:equalfreq} to \ref{sec:jordanblock} and is shown to be described by Jordan blocks.

\section{Eigenfunctions of  Oscillator Hamiltonian $H_0$} \label{shoham}
The  Euclidean version of the acceleration Hamiltonian \cite{mannheimmain}  is given by
\begin{equation}
\label{mainham}
H=-\frac{1}{2\gamma}\frac{\partial^2}{\partial v^2}-v\frac{\partial}{\partial x}+\frac{\gamma}{2}(\omega_1^2+\omega_2^2)v^2+\frac{\gamma}{2}\omega_1^2\omega_2^2x^2
\end{equation}
Note $H$ is a symmetric function of $\omega_1,\omega_2$; we choose $\omega_1 >\omega_2$.

The equivalent Hermitian Hamiltonian $H_O$ is given by the following
\begin{align}
\label{eq:rep4}
H&=e^{Q/2}H_{O}~e^{-Q/2}~~;~~H^\dagger=e^{-Q/2}H_{O}~e^{Q/2}\\
\label{eq:rep3}
H_O&=-\frac{1}{2\gamma}\frac{\partial^2}{\partial v^2}-\frac{1}{2\gamma\omega_1^2}\frac{\partial^2}{\partial x^2}+\frac{\gamma}{2}\omega_1^2v^2+\frac{\gamma}{2}\omega_1^2 \omega_2^2x^2
\end{align}
where the similarity transformation
\begin{align}
\label{Qoperdef}
Q&=axv-b\frac{\partial^2}{\partial x \partial v}
\end{align}
has coefficients, from Paper I, given for $\omega_1 >\omega_2$ by
\begin{equation}
\label{abqparamaet}
\sqrt{\frac{a}{b}}=\gamma\omega_1\omega_2;~~\sinh(\sqrt{ab})=\frac{2\omega_1\omega_2}{\omega_1^2-\omega_2^2}~~\Rightarrow~~\sqrt{ab}=\ln\left(\frac{\omega_1+\omega_2}{\omega_1-\omega_2}\right)
\end{equation}

The Hamiltonian $H$ is pseudo-Hermitian as long as there is a well defined similarity transformation $Q$. The similarity transformation becomes singular for $\omega_1 =\omega_2$ and $H$ becomes inequivalent to a Hermitian Hamiltonian; this case is studied in detail in Sections \ref{sec:equalfreq} to \ref{sec:jordanblock}. 

The Euclidean Hamiltonian has two distinct and independent state spaces', namely the state space $\mathcal{V}$ of the non-Hermitian Hamiltonian  $H$ and  the state space of the Hermitian Hamiltonian  $H_O$, namely $\mathcal{V}_O$ \cite{mannheim1}, \cite{mannheimmain}. 
 
Both the oscillator state space $\mathcal{V}_O$ and the acceleration state space $\mathcal{V}$ have a completeness equation given by
\begin{equation}
\mathbb{I}=\int dxdv \ket{x,v}\bra{x,v}
\end{equation}

The oscillator structure of $H_0$ yields two sets of creation and destruction operators, given by the following
\begin{align*}
a_v&=\sqrt{\frac{\gamma\omega_1}{2}}\left[v+\frac{1}{\gamma\omega_1}\frac{\partial}{\partial v}\right]~~;~~a_v^\dagger=\sqrt{\frac{\gamma\omega_1}{2}}\left[v-\frac{1}{\gamma\omega_1}\frac{\partial}{\partial v}\right]\\
a_x&=\sqrt{\frac{\gamma\omega_1^2\omega_2}{2}}\left[x+\frac{1}{\gamma\omega_1^2\omega_2}\frac{\partial}{\partial x}\right]~~;~~a_x^\dagger=\sqrt{\frac{\gamma\omega_1^2\omega_2}{2}}\left[x-\frac{1}{\gamma\omega_1^2\omega_2}\frac{\partial}{\partial x}\right]\\
&\Rightarrow [a_v,a_v^\dagger]=1= [a_x,a_x^\dagger]~~; ~~[a_x,a_v^\dagger] =0 = [a_x,a_v] 
\end{align*}
The creation operators $a_v^\dagger,~a_x^\dagger$ are the Hermitian conjugate of the destruction operators $a_v,~a_x$.

From above
\begin{align}
v&=\sqrt{\frac{1}{2\gamma\omega_1}}(a_v+a_v^\dagger)~~;~~
\frac{\partial}{\partial v}=\sqrt{\frac{\gamma \omega_1}{2}}(a_v-a_v^\dagger)\\
x&=\sqrt{\frac{1}{2\gamma\omega_1^2\omega_2}}(a_x+a_x^\dagger)~~;~~
\frac{\partial}{\partial x}=\sqrt{\frac{\gamma\omega_1^2\omega_2}{2}}(a_x-a_x^\dagger)\nonumber
\end{align}

The oscillator Hamiltonian, from Eq. \ref{eq:rep3}, is given by
\begin{equation}
H_{O}=\omega_1a_v^\dagger a_v+\omega_2a_x^\dagger a_x+\frac{1}{2}(\omega_1+\omega_2)
\end{equation}
The vacuum is defined, as usual, by requiring that there be no excitations, namely
\begin{eqnarray*}
&&a_v\ket{0,0}=a_x\ket{0,0}=0\\
&&H_O\ket{0,0}=E_0\ket{0,0}~~;~~E_0=\frac{1}{2}(\omega_1+\omega_2)
\end{eqnarray*}
The coordinate representation of the oscillator vacuum state is given by
\begin{eqnarray*}
&&\langle x,v\ket{0,0}=\left(\frac{\gamma^2\omega_1^3\omega_2}{\pi^2}\right)^{1/4}\exp\{-\frac{1}{2}[\gamma\omega_1 v^2+\gamma\omega_1^2\omega_2 x^2]\}
\end{eqnarray*}

The energy eigenfunctions $\ket{n,m}$ and the dual eigenfunctions $\bra{n,m}$  have eigenenergies $E_{nm}$ are given by
\begin{eqnarray}
&&H_0\ket{n,m}=E_{nm}\ket{n,m} ~~;~~\bra{n,m} H_0=E_{nm}\bra{n,m} \nonumber\\
\label{eigenfunctionmn}
&&\ket{n,m}=\frac{a_v^{\dagger n}}{\sqrt{n!}}\frac{a_x^{\dagger  m}}{\sqrt{m!}}\ket{0,0}~;~\bra{x,v}m,n\rangle=\langle m,n\ket{x,v}:~\text{real}\\
\label{energymn}
&&E_{nm}=n\omega_1+m\omega_2+E_0=n\omega_1+m\omega_2+\frac{1}{2}(\omega_1+\omega_2)
\end{eqnarray}
satisfy the orthonormality equation
\begin{equation}
\bra{n',m'}n,m\rangle=\delta_{n-n'}\delta_{m-m'}
\end{equation}
Hence, the spectral representation of $H_O$ is given by
\begin{equation}
\label{hzerospec}
H_O=\sum_{mn}E_{mn}\ket{m,n}\bra{m,n}
\end{equation}

All the state vectors for $H_0$ have been defined entirely on state space $\mathcal{V}_O$ with no reference to state space $\mathcal{V}$ of Euclidean Hamiltonian $H$. 

\section{Eigenfunctions of $H$ and $H^\dagger$} \label{hamacclnact}
The left and right eigenfunctions of $H$ are different since $H$ is non-Hermitian; denote the right \textit{eigenfunctions}  by $\ket{\Psi_{mn}}$ and left \textit{dual eigenfunctions} by  $\bra{\Psi^D_{mn}}$. The notation $\bra{\Psi^D_{mn}}$ denotes the dual of the state $\ket{\Psi_{mn}}$ to differentiate it from the state $\bra{\Psi_{mn}}$ that is obtained from $\ket{\Psi_{mn}}$ by complex conjugation. 

The energy eigenfunctions of Hamiltonian $H$, from Eq.~\ref{eq:rep4} are given by 
Eq.~\ref{eq:rep3} yields
\begin{align}
H\ket{\Psi_{mn}}&=E_{nm}\ket{\Psi_{mn}}~~;~~
\ket{\Psi_{mn}}=e^{Q/2}\ket{m,n}\nonumber\\
\bra{\Psi^D_{mn}}H&=E_{nm}\bra{\Psi^D_{mn}}~~;~~
\bra{\Psi^D_{mn}}=\bra{m,n}e^{-Q/2}=\bra{\Psi_{mn}}e^{-Q}
\end{align}
Since the eigenfunctions of $H_0$ are complete, the completeness equation for the Hilbert space on which the pseudo-Hermitian Hamiltonian $H$ acts on is given by
\begin{align}
\label{completeeqn}
\mathbb{I}=\sum_{m,n=1}^\infty  \ket{\Psi_{mn}} \bra{\Psi^D_{mn}} =\sum_{m,n=1}^\infty  \ket{\Psi_{mn}} \bra{\Psi_{mn}} e^{-Q}
\end{align}
For many of the computations, it is convenient to separate out the overall normalization constants of the eigenfunctions. Define
\begin{align}
\label{normmn}
&~~~~~~~~~~~~\ket{\Psi_{mn}} =N_{mn} \ket{\psi_{mn}}~;~~\bra{\Psi^D_{mn}} =N_{mn} \bra{\psi^D_{mn}}\nonumber\\
&\bra{\Psi^D_{mn}}\Psi_{mn}\rangle=N_{mn}^2 \bra{\psi^D_{mn}}\psi_{mn}\rangle=1~;~~N_{mn}=N^*_{mn}>0:~\text{real;~positive}
\end{align}
All the states $\ket{m,n}$ are real, as are all the matrix elements of $e^{Q/2}$ given in Paper I; hence the coordinate representations of $\ket{\Psi_{mn}}$ are real and given by
\begin{align}
\Psi_{mn}(x,v)=\langle x,v\ket{\Psi_{mn}}=\Psi^*_{mn}(x,v):~\text{real}
\end{align}

It can be directly verified by applying $H$ given in Eq. \ref{mainham}, that the vacuum state and its dual are given by
\begin{align}
\label{vacuumhigh}
&\Psi_{00}(x,v)=\langle x,v\ket{\Psi_{00}}=\bra{x,v}e^{Q/2}\ket{0,0}\nonumber\\
&~~~=N_{00}\exp\left\{-\frac{\gamma}{2}(\omega_1+\omega_2)\omega_1\omega_2x^2-\frac{\gamma}{2}(\omega_1+\omega_2)v^2-\gamma\omega_1\omega_2xv \right\}\\
\label{vacqderivdual} 
&\Psi_{00}^D(x,v)=\bra{\Psi_{00}^D}x,v\rangle=\bra{0,0}e^{-Q/2}\ket{x,v} =\bra{x,v}e^{-Q/2}\ket{0,0}\nonumber\\
&=N_{00}\exp\{-\frac{\gamma}{2}(\omega_1+\omega_2)\omega_1\omega_2x^2-\frac{\gamma}{2}(\omega_1+\omega_2)v^2+\gamma\omega_1\omega_2xv\}\\
\label{vacqderivnorm} 
&\Rightarrow \bra{\Psi_{00}^D} \Psi_{00}\rangle=1~~\Rightarrow~~N_{00}=(\omega_1 \omega_2)^{1/4} \sqrt{\frac{\gamma}{\pi}(\omega_1+\omega_2)}
\end{align}

One can write the completeness equation in the following manner
\begin{align}
\label{completeeqn7}
\mathbb{I}=\sum_{m,n=1}^\infty  \ket{\psi_{mn}}N^2_{mn} \bra{\psi^D_{mn}} =\sum_{m,n=1}^\infty  \ket{\psi_{mn}}N^2_{mn} \bra{\psi_{mn}} e^{-Q}
\end{align}
The completeness equation above in Eq. \ref{completeeqn7} reflects one major difference between the Euclidean and Minkowski formulation of this model. In the Minkowski case, for $\omega_1>\omega_2$, the Minkowski  normalization constants $\tilde{N}_{mn}$ (differing by $\pm$ signs from the corresponding Euclidean coefficients) can be both \textit{negative} or \textit{positive}. Extra coefficients, called $c_n$ by Bender and Mannheim $ \cite{mannheim1}$,  need to introduced in the completeness equation by hand, \textit{in addition} to $\tilde{N}_{mn}$ -- to compensate for the coefficients  $\tilde{N}_{mn}$ that have a negative sign. 

The  Minkowski velocity is transformed to a pure imaginary Euclidean velocity and this leads to a state space for the Euclidean Hamiltonian that has all positive norm eigenstates. There is no need for any extra coefficients in the Euclidean formulation and the Minkowski coefficients $c_n$ can be recovered from the Euclidean state space by analytically continuing back to Minkowski time.  The Euclidean completeness equation is completely determined by  positive norm state vectors.

From Eqs. \ref{eq:rep3} and \ref{hzerospec}, the Hamiltonian $H$ has the following spectral decomposition
\begin{align}
\bra{x,v}H\ket{x',v'}&=\bra{x,v}e^{Q/2}H_{O}~e^{-Q/2}\ket{x',v'}\\
   &=\sum_{mn}E_{mn}\langle x,v|e^{Q/2}\ket{m,n}\bra{m,n}e^{-Q/2}\ket{x',v'}\\
   &=\sum_{mn}E_{mn}\Psi_{mn}(x,v)\Psi^D_{mn}(x',v')
\end{align}

As mentioned earlier, unlike the case for a Hermitian Hamiltonian, for the non-Hermitian case the dual eigenfunction $\bra{\Psi^D}$ is not the complex conjugate of $\ket{\Psi}$. The dual eigenstates $\Psi^D_{mn}(x,v)$ are given by
\begin{align}
&\bra{\Psi^D_{mn}}=\bra{m,n}e^{-Q/2}\\
&\Psi^D_{mn}(x,v)=\bra{\Psi^D_{mn}} x,v\rangle=\bra{m,n}e^{-Q/2}|x,v\rangle
\end{align}

To obtain the matrix element of a non-Hermitian operator, note that all operators are defined by their action \textit{only} on the dual space. This fact is un-important for Hermitian operators since acting on the state space or its dual space is equivalent but this is \textit{not} true for non-Hermitian operators, since the result depends on which space the operators act on. 

The matrix elements of operator $\hat{O}(x,v,\partial_x,\partial_v)$ is defined by the following
\begin{equation}
\bra{x,v}\hat{O}\ket{\Psi}=\hat{O}(x,v,\partial_x,\partial_v) \Psi(x,v)
\end{equation}
The matrix element of the operator $\hat{O}(x,v,\partial_x,\partial_v)$ acting on the dual state vector $\langle \Psi|$ is given in terms of the conjugate operator $\hat{O}^\dagger(x,v,\partial_x,\partial_v)$ as follows
\begin{align}
\label{conjuope}
\bra{\Psi}\mathcal O \ket{x,v}^*&\equiv \bra{x,v}\mathcal O^\dagger\ket{\Psi} =\mathcal O^\dagger(x,v,\partial_x,\partial_v)\Psi(x,v)
\end{align}
Hence, the Hamiltonian $H$ has the following spectral decomposition
\begin{align}
\label{specfinal}
\bra{x,v}H\ket{x',v'} &=\sum_{mn}E_{mn}\Psi_{mn}(x,v)\Psi^D_{mn}(x',v')
\end{align}

Since $\Psi^D_{mn}=(\Psi^D_{mn})^*$, the matrix element $\bra{\Psi^D_{mn}}H\ket{x,v}$ of the dual eigenfunction  $\bra{\Psi^D_{mn}}$ is given as follows
\begin{align}
\bra{\Psi^D_{mn}}H\ket{x,v}^*&=\bra{x,v}H^\dagger \ket{\Psi^D_{mn}}=H^\dagger \Psi^D_{mn}(x,v)
\end{align}
Note from the general form of the Hamiltonian $H$, we have that
\begin{align}
H^\dagger (x,v,\partial/\partial v,\partial/\partial)=H (x,-v,-\partial/\partial v,\partial/\partial x)\\
                                               =H (-x,v,\partial/\partial v,-\partial/\partial x)
\end{align}
The only difference between $\Psi^D_{mn}(x,v)$ and $\Psi_{mn}(x,v)$ is that $v \to -v$ \textit{or} $x \to -x$. Since all the eigenfunctions are real, one obtains the general result that
\begin {eqnarray}
\label{thetafn}
\Psi^D_{mn}(x,v) =  
\left\{ \begin{array}{ll}
\Psi_{mn}(x,-v) &  \\
\text{or} &  \\
\Psi_{mn}(-x,v) &
\end{array}
\right.    
\end {eqnarray}
and it follows that
\begin{align*}
 H^\dagger \Psi^D_{mn}(x,v)&=H^\dagger \Psi_{mn}(x,-v)=E_{mn}\Psi_{mn}(x,-v)\\
\text{or}\\
 H^\dagger \Psi^D_{mn}(x,v)&=H^\dagger \Psi_{mn}(-x,v)=E_{mn}\Psi_{mn}(-x,v)
\end{align*}

The rules for conjugation choose the dual eigenvectors precisely in a manner that guarantees that $\bra{\Psi^D_{mn}}\Psi_{mn}\rangle=1$ as in Eq. \ref{normmn} and yields a positive norm (Hilbert) state space for the pseudo-Hermitian Hamiltonian $H$.

\section{Excited states of $H$} \label{excitedstham}
Recall from Eq. \ref{mainham} that the Hamiltonian is given by
\begin{equation}
\label{hamw1w2}
H=-\frac{1}{2\gamma}\frac{\partial^2}{\partial v^2}-v\frac{\partial}{\partial x}+\frac{\gamma}{2}(\omega_1^2+\omega_2^2)v^2+\frac{\gamma}{2}\omega_1^2\omega_2^2x^2
\end{equation}
and from Eq. \ref{eq:rep3}
\begin{align*}
%\label{eq:rep9}
e^{-Q/2}He^{Q/2}&=H_{O}\\
%\label{eq:rep3}
&=-\frac{1}{2\gamma}\frac{\partial^2}{\partial v^2}-\frac{1}{2\gamma\omega_1^2}\frac{\partial^2}{\partial x^2}+\frac{\gamma}{2}\omega_1^2v^2+\frac{\gamma}{2}\omega_1^2 \omega_2^2x^2
\end{align*}

To illustrate the general features of the eigenfunctions, the first few eigenfunctions are evaluated. Recall that
\begin{align}
\label{eq:adag1}
&a_v^\dagger=\sqrt{\frac{\gamma\omega_1}{2}}\left(v-\frac{1}{\gamma\omega_1}\frac{\partial }{\partial v}\right)\\
\label{eq:agag2}
&a_x^\dagger=\sqrt{\frac{\gamma\omega_1^2\omega_2}{2}}\left(x-\frac{1}{\gamma\omega_1^2\omega_2}\frac{\partial }{\partial x}\right)
\end{align}

In general, one can find the explicit co-ordinate representation of any eigenfunction by the following procedure
\begin{align}
\Psi_{nm}(x,v)&=\bra{x,v}e^{Q/2}\ket{m,n}=\bra{x,v}e^{Q/2}\left\{\frac{a_v^{\dagger n}}{\sqrt{n!}}\frac{a_x^{\dagger m}}{\sqrt{m!}}\right\}\ket{0,0}\nonumber\\
&=\bra{x,v}e^{Q/2}\frac{a_v^{\dagger n}}{\sqrt{n!}}\frac{a_x^{\dagger m}}{\sqrt{m!}}e^{-Q/2}e^{Q/2}\ket{0,0}\nonumber\\
&=e^{Q/2}\frac{a_v^{\dagger n}}{\sqrt{n!}}\frac{a_x^{\dagger m}}{\sqrt{m!}}e^{-Q/2}\Psi_{00}(x,v)
\end{align}
where, using Eqs. \ref{a1transfor} and \ref{a1transfor2} given below, one can explicitly evaluate
\begin{align}
&e^{Q/2}a_v^{\dagger n}a_x^{\dagger m}e^{-Q/2}=\left(e^{Q/2}a_v^\dagger e^{-Q/2}\right)^n\left(e^{Q/2}a_x^\dagger e^{-Q/2}\right)^m 
\label{eq:a2}
\end{align}

\subsection{Energy $\omega_1$ eigenstate $\Psi_{10}(x,v)$}
The single  $\omega_1$ excitation energy eigenstate state is the following
\begin{align}
\Psi_{10}(x,v)&=\bra{x,v}e^{Q/2}\Big\{a_v^\dagger \ket{0,0}\Big\} =\bra{x,v}e^{Q/2}a_v^\dagger e^{-Q/2}e^{Q/2}\ket{0,0} \nonumber\\
   \label{onezeroiniti}
 \Rightarrow \Psi_{10}(x,v)  &=e^{Q/2}a_v^\dagger e^{-Q/2}\Psi_{00}(x,v)
   %   &=\left[Av-\frac{BC}{\gamma \omega_1}x +\frac{B}{C}\frac{\partial}{\partial x}-\frac{A}{\gamma \omega_1}\frac{\partial}{\partial v}\right]\Psi_{00}(x,v)
\end{align}

The fundamental similarity transformation given in Eq. \ref{Qoperdef} yields, using the results from Paper I, the following
\begin{align}
%\label{eq:rep0}
e^{Q/2}ve^{-Q/2}&=Av-\frac{B}{C}\frac{\partial }{\partial x}\nonumber\\
e^{Q/2}\frac{\partial}{\partial v}e^{-Q/2}&=A\frac{\partial }{\partial v}-BCx
\end{align}
with  the coefficients functions being given by
\begin{align}
\label{coeffab}
A=\frac{\omega_1}{\sqrt{\omega_1^2-\omega_2^2}}~~;~~B=\frac{\omega_2}{\sqrt{\omega_1^2-\omega_2^2}}~~;~~C=\gamma\omega_1\omega_2 
\end{align}

Hence, from Eq. \ref{eq:adag1}
\begin{align}
e^{Q/2}a_v^\dagger e^{-Q/2}&=\sqrt{\frac{\gamma\omega_1}{2}}e^{Q/2}\left[v-\frac{1}{\gamma\omega_1}\frac{\partial}{\partial v}\right]e^{-Q/2} \nonumber\\
&=\sqrt{\frac{\gamma\omega_1}{2}}\left[Av+\frac{BC}{\gamma \omega_1}x -\frac{B}{C}\frac{\partial}{\partial x}-\frac{A}{\gamma \omega_1}\frac{\partial}{\partial v}\right]\nonumber\\
\label{a1transfor}
\Rightarrow e^{Q/2}a_v^\dagger e^{-Q/2}&=\sqrt{\frac{\gamma\omega_1}{2(\omega_1^2-\omega_2^2)}}\left[\omega_1v+ \omega_2^2 x-\frac{1}{\gamma\omega_1}\frac{\partial}{\partial x} -\frac{1}{\gamma}\frac{\partial}{\partial v} \right]
%&=\sqrt{\frac{2\gamma\omega_1}{\omega_1^2-\omega_2^2}}(\omega_1 +\omega_2)\left[v+ \omega_2 x \right]\Psi_{00}(x,v)
\end{align}

Hence, from Eqs. \ref{onezeroiniti} and \ref{a1transfor}
\begin{align*}
\Psi_{10}(x,v)
%&=\sqrt{\frac{\gamma\omega_1}{2}}\left[Av+\frac{BC}{\gamma \omega_1}x -\frac{B}{C}\frac{\partial}{\partial x}-\frac{A}{\gamma \omega_1}\frac{\partial}{\partial v}\right]\Psi_{00}(x,v)\\
&=\sqrt{\frac{\gamma\omega_1}{2(\omega_1^2-\omega_2^2)}}\left[\omega_1v+ \omega_2^2 x-\frac{1}{\gamma\omega_1}\frac{\partial}{\partial x} -\frac{1}{\gamma}\frac{\partial}{\partial v} \right]\Psi_{00}(x,v)
\end{align*}
Using the explicit representation of the vacuum state $\Psi_{00}(x,v)$ given in Eq. \ref{vacuumhigh} yields the final result
\begin{align}
\label{velfirsteigen}
\Psi_{10}(x,v)&=\sqrt{\frac{2\gamma\omega_1}{\omega_1^2-\omega_2^2}}(\omega_1 +\omega_2)\left[v+ \omega_2 x \right]\Psi_{00}(x,v)
\end{align}

The dual energy eigenstate is defined by
\begin{align}
&\langle\Psi_{10}^D\ket{x,v}=\bra{1,0}e^{-Q/2}\ket{x,v} =\bra{0,0}a_ve^{-Q/2}\ket{x,v} \nonumber
\end{align}
Using the rule for conjugate operators given in Eq. \ref{conjuope} yields
\begin{align}
&\langle\Psi_{10}^D\ket{x,v}^* =\bra{x,v}e^{-Q/2}a_v^\dagger\ket{0,0}=\bra{x,v}e^{-Q/2}a_v^\dagger e^{Q/2}e^{-Q/2}\ket{0,0}\nonumber\\
\label{dualonezero}
&\Rightarrow  \Psi_{10}^D (x,v)= \bra{x,v}\Psi_{10}^D \rangle =   e^{-Q/2}a_v^\dagger e^{Q/2} \Psi_{00}^D(x,v)                      
\end{align}
Note all the eigenfunctions are real, and in particular $\bra{x,v}\Psi_{10}^D \rangle=\langle \Psi_{10}^D \ket{x,v}=\Psi_{10}^D (x,v)$ and  $\bra{x,v}\Psi_{00}^D \rangle=\langle \Psi_{00}^D \ket{x,v}=\Psi_{10}^D (x,v)$. 

Similar to the derivation of Eq. \ref{a1transfor},  one obtains the following
\begin{align}
\label{a1transforadjoint}
e^{-Q/2}a_v^\dagger e^{Q/2}&=\sqrt{\frac{\gamma\omega_1}{2(\omega_1^2-\omega_2^2)}}\left[\omega_1v - 
\omega_2^2 x+\frac{1}{\gamma\omega_1}\frac{\partial}{\partial x} -\frac{1}{\gamma}\frac{\partial}{\partial v} \right]
%&=\sqrt{\frac{2\gamma\omega_1}{\omega_1^2-\omega_2^2}}(\omega_1 +\omega_2)\left[v+ \omega_2 x \right]\Psi_{00}(x,v)
\end{align}
and, from Eq. \ref{dualonezero}, yields the dual eigenfunction
\begin{align*}
\Psi^D_{10}(x,v)&=\sqrt{\frac{\gamma\omega_1}{2(\omega_1^2-\omega_2^2)}}\left[\omega_1v - 
\omega_2^2 x+\frac{1}{\gamma\omega_1}\frac{\partial}{\partial x} -\frac{1}{\gamma}\frac{\partial}{\partial v} \right]\Psi^D_{00}(x,v)
\end{align*}
Using the explicit representation of the vacuum state $\Psi^D_{00}(x,v)$ given in Eq. \ref{vacqderivdual} yields the final result
\begin{align}
\label{velfirsteigendual}
\Psi^D_{10}(x,v)&=\sqrt{\frac{2\gamma\omega_1}{\omega_1^2-\omega_2^2}}(\omega_1 +\omega_2)\left[v- \omega_2 x \right]\Psi^D_{00}(x,v)
\end{align}

\subsection{Energy $\omega_2$ eigenstate $\Psi_{01}(x,v)$}
The one $\omega_2$  excitation energy eigenstate is given by the following.
\begin{align}
\Psi_{01}(x,v)&=\bra{x,v}e^{Q/2}\Big\{a_x^\dagger \ket{0,0}\Big\}=\bra{x,v}e^{Q/2}a_x^\dagger e^{-Q/2}e^{Q/2}\ket{0,0}\nonumber\\
\label{secondinitial}
\Rightarrow  \Psi_{01}(x,v)  &=e^{Q/2}a_x^\dagger e^{-Q/2}\Psi_{00}(x,v)
   %&=\left[Ax-\frac{BC}{\gamma \omega_1^2\omega_2}v +\frac{B}{C}\frac{\partial}{\partial v}-\frac{A}{\gamma \omega_1^2\omega_2}\frac{\partial}{\partial x}\right]\Psi_{00}(x,v)
\end{align}

The similarity transformation given in Paper I yields the following
\begin{align}
\label{eq:commu}
%e^{-Q}xe^{ \tau Q}&=\cosh(\sqrt{ab})x+\sqrt{\frac{b}{a}}\sinh(\sqrt{ab})\frac{\partial }{\partial v}\nonumber\\
%e^{-Q}\frac{\partial}{\partial x}e^{ \tau Q}&=\cosh(\sqrt{ab})\frac{\partial }{\partial x}+\sqrt{\frac{a}{b}}\sinh(\sqrt{ab})v\nonumber\\
%\label{eq:rep0}
e^{Q/2}xe^{-Q/2}&=Ax-\frac{B}{C}\frac{\partial }{\partial v}\nonumber\\
e^{Q/2}\frac{\partial}{\partial x}e^{-Q/2}&=A\frac{\partial }{\partial x}-B C v
\end{align}

Hence, from Eq. \ref{eq:agag2}
\begin{align}
e^{Q/2}a_x^\dagger e^{-Q/2}&=\sqrt{\frac{\gamma\omega_1^2\omega_2}{2}}e^{Q/2}\left[x-\frac{1}{\gamma\omega_1^2\omega_2}\frac{\partial }{\partial x}\right] e^{-Q/2} \nonumber\\
&=\sqrt{\frac{\gamma\omega_1^2\omega_2}{2}}\left[Ax+\frac{BC}{\gamma \omega_1^2\omega_2}v -\frac{B}{C}\frac{\partial}{\partial v}-\frac{A}{\gamma \omega_1^2\omega_2}\frac{\partial}{\partial x}\right]\nonumber\\
\label{a1transfor2}
\Rightarrow e^{Q/2}a_x^\dagger e^{-Q/2}&=\sqrt{\frac{\gamma\omega_2}{2(\omega_1^2-\omega_2^2)}}\left[\omega_2v+\omega_1^2 x-\frac{1}{\gamma\omega_2}\frac{\partial}{\partial x} -\frac{1}{\gamma}\frac{\partial}{\partial v} \right]
%&=\sqrt{\frac{2\gamma\omega_1}{\omega_1^2-\omega_2^2}}(\omega_1 +\omega_2)\left[v+ \omega_2 x \right]\Psi_{00}(x,v)
\end{align}

Hence, from Eqs. \ref{secondinitial} and \ref{a1transfor2}
\begin{align*}
\Psi_{01}(x,v)
%&=\sqrt{\frac{\gamma\omega_1}{2}}\left[Av+\frac{BC}{\gamma \omega_1}x -\frac{B}{C}\frac{\partial}{\partial x}-\frac{A}{\gamma \omega_1}\frac{\partial}{\partial v}\right]\Psi_{00}(x,v)\\
&=\sqrt{\frac{\gamma\omega_2}{2(\omega_1^2-\omega_2^2)}}\left[\omega_2v+\omega_1^2 x-\frac{1}{\gamma\omega_2}\frac{\partial}{\partial x} -\frac{1}{\gamma}\frac{\partial}{\partial v} \right]\Psi_{00}(x,v)
\end{align*}
Using the explicit representation of the vacuum state $\Psi_{00}(x,v)$ given in Eq. \ref{vacuumhigh} yields the final result
\begin{align}
\label{firsxeigenst}
\Psi_{01}(x,v) &=\sqrt{\frac{2\gamma\omega_2}{\omega_1^2-\omega_2^2}}(\omega_1 +\omega_2)\left[v+ \omega_1 x \right]\Psi_{00}(x,v)
\end{align}

The dual energy eigenstate, similar to Eq. \ref{dualonezero}, is defined by
\begin{align}
\label{dualzeroone}
 \Psi_{01}^D(x,v) &=   e^{-Q/2}a_v^\dagger e^{Q/2}e^{-Q/2} \Psi_{00}^D(x,v)                      
\end{align}

Eqs. \ref{eq:agag2} and \ref{eq:commu} yield
\begin{align}
\label{a1transforadjoint2}
e^{-Q/2}a_x^\dagger e^{Q/2}&=\sqrt{\frac{\gamma\omega_2}{2(\omega_1^2-\omega_2^2)}}\left[-\omega_2v+\omega_1^2 x-\frac{1}{\gamma\omega_2}\frac{\partial}{\partial x} +\frac{1}{\gamma}\frac{\partial}{\partial v} \right]
%&=\sqrt{\frac{2\gamma\omega_1}{\omega_1^2-\omega_2^2}}(\omega_1 +\omega_2)\left[v+ \omega_2 x \right]\Psi_{00}(x,v)
\end{align}
Using the explicit representation of the vacuum state $\Psi^D_{00}(x,v)$ given in Eq. \ref{vacqderivdual} and from Eq. \ref{dualzeroone} yields the dual eigenfunction
\begin{align}
\label{firsxeigenstdual}
\Psi^D_{01}(x,v) &=\sqrt{\frac{2\gamma\omega_2}{\omega_1^2-\omega_2^2}}(\omega_1 +\omega_2)\left[-v+ \omega_1 x \right]\Psi^D_{00}(x,v)
\end{align}

We collect  the results for the first two one excitation states; the normalization constants are separated out for later convenience; using the value of $N_{00}$ given in Eq. \ref{vacqderivnorm}  yields the following
\begin{align}
\label{firsttwostatessummary}
&\Psi_{10}(x,v) =N_{10}\left[v+ \omega_2 x \right]\psi_{00}(x,v)~;~
\Psi^D_{10}(x,v) =N_{10}\left[v- \omega_2 x \right]\psi^D_{00}(x,v) \nonumber\\
&\Psi_{01}(x,v) =N_{01}\left[v+ \omega_1 x \right]\psi_{00}(x,v)~;~
\Psi^D_{01}(x,v) =N_{01}\left[-v+ \omega_1 x \right]\psi^D_{00}(x,v)\nonumber\\
&N_{10}=\gamma\sqrt{2}\frac{(\omega_1+\omega_2)}{\sqrt{\pi(\omega_1-\omega_2)}}~\omega_1^{3/4}\omega_2^{1/4}~;~N_{01}=\gamma\sqrt{2}\frac{(\omega_1+\omega_2)}{\sqrt{\pi(\omega_1-\omega_2)}}~\omega_1^{1/4}\omega_2^{3/4}\nonumber\\
&E_{10}=\omega_1+E_{00}~~;~~E_{01}=\omega_2+E_{00}
\end{align}
The eigenstates are orthogonal and normalized, namely
\begin{align*}
%\label{firsttwostatessummary}
&\langle\Psi^D_{10}|\Psi_{10}\rangle=1=  \langle\Psi^D_{01}|\Psi_{01}\rangle~~;~~\langle\Psi^D_{10}|\Psi_{01}\rangle=0
\end{align*}

Note the remarkable result that under a duality transformation, the dual eigenstates have a transformation that depends on the eigenstate; in particular
\begin{align}
\label{firsttwostatdual}
&\Psi^D_{10}(x,v)=\Psi_{10}(-x,v)  \nonumber\\
&\Psi^D_{01}(x,v)=\Psi_{01}(x,-v) 
\end{align} 
This feature generalizes to all the energy eigenstates and guarantees that the state space, for $\omega_1>\omega_2$,  always has a positive norm.
 
The first two energy eigenstates of the pseudo-Hermitian Hamiltonian $H$ are the first excitation of the position degree of freedom $x$, given by $\Psi_{01}$ and the velocity degree of freedom $v$, given by $\Psi_{10}$. In the limit of $\omega_1 \to \omega_2$, the eigenstates $\Psi_{01}$ and  $\Psi_{01}$ become degenerate. This important property of the energy eigenspectrum will be studied in some detail when the limit of $\omega_1 \to \omega_2$ is Section \ref{sec:equalfreq}.

\section{Propagator: $\omega_1 > \omega_2$} \label{sec:proppseudo}
The infinite time path integral is given by
\begin{align*}
Z&=\lim_{T\to\infty}tr\left(e^{-TH}\right)=
\int Dx e^{\mathcal{S}}\\
\mathcal{S}&=-\frac{1}{2}\gamma\int_{-\infty}^{+\infty}dt\left[\ddot{x}^2+(\omega_1^2+\omega_2^2)\dot{x}^2+\omega_1^2\omega_2^2{x}^2\right]
\end{align*}
The propagator is given by the path integral
\begin{align}
G(\tau)&=\frac{1}{Z}\int Dx e^{\mathcal{S}}x(t)x(t')
%\mathcal{S}&=-\frac{1}{2}\gamma\int_{-\infty}^{+\infty}dt\left[\ddot{x}^2+(\omega_1^2+\omega_2^2)+\dot{x}^2+\omega_1^2\omega_2^2{x}^2\right]
\end{align}
The acceleration action is a quadratic functional of the paths $x(t)$ and can be evaluated exactly. Define the Fourier transformed variables that diagonalize the action, namely
\begin{align}
x(t)&=\int_{-\infty}^{+\infty}\frac{dk}{2\pi}e^{ikx}x_k\\
\Rightarrow S&=-\frac{1}{2}\gamma\int_{-\infty}^{+\infty}dk\left[k^4+(\omega_1^2+\omega_2^2)k^2+\omega_1^2\omega_2^2\right]x_{-k}x_k
\end{align}
Using Gaussian path integration yields
\begin{align}
G(\tau)&=\frac{1}{\gamma}\int_{-\infty}^{+\infty}\frac{dk}{2\pi}\frac{e^{ik(t-t')}}{(k^2+\omega_1^2)(k^2+\omega_2^2)}\nonumber\\
\label{propomega}
&=\frac{1}{2\gamma}\frac{1}{\omega_1^2-\omega_2^2}\left[\frac{e^{-\omega_2\tau}}{\omega_2}-\frac{e^{-\omega_1\tau}}{\omega_1}\right]~~;~~\tau=|t-t'|
\end{align}
where the last equation has been obtained using counter integration.% that Eq. \ref{propomega} yields the following two branches for the propagator.

Constructing the propagator by inserting the complete set of states yields a realization of the propagator in terms of the state space and Hamiltonian. The state space definition of the propagator is given by
\begin{align*}
G(\tau)&=\lim_{T\to\infty}\frac{1}{Z} tr\left(e^{-(T-\tau)H}xe^{-\tau H} x\right)~~;~~\tau=|t-t'|
\end{align*}
Note that
\begin{align*}
&\lim_{T\to\infty} e^{-TH}\simeq e^{-TE_0}\ket{\Psi_{00}}\bra{\Psi_{00}}e^{-Q}=e^{-TE_0}e^{Q/2}\ket{0,0}\bra{0,0}e^{-Q/2}\\
&tr(e^{-TH})\simeq e^{-TE_0}
\end{align*}

Since
\begin{align*}
H&=e^{Q/2}H_{O}e^{-Q/2}~~;~~ \ket{\Psi_{00}}=e^{Q/2}\ket{0,0}~~;~~\bra{\Psi_{00}}=\bra{0,0}e^{Q/2}
\end{align*}
the propagator is given by
\begin{align}
G(\tau)&=\lim_{T\to\infty}\frac{1}{Z} tr\left(e^{-(T-\tau)H}xe^{-\tau H} x\right)~~;~~\tau=|t-t'|\nonumber\\
\label{proeqnvev2}
&=\bra{\Psi_{00}}e^{-Q}xe^{-\tau (H-E_0)}x\ket{\Psi_{00}}\\
\label{propmatele}
   &=\bra{0,0}e^{-Q/2}xe^{Q/2}e^{-\tau (H_0-E_0)}e^{-Q/2}xe^{Q/2}\ket{0,0}
\end{align}
%Note Eq. \ref{proeqnvev2} has been obtained earlier in Eq. \ref{proeqnvev} based on a Hilbert space derivation of the propagator.

From Paper I
\begin{align}
\label{eqxemq}
e^{-Q/2}xe^{Q/2}&=Ax+BC\frac{\partial}{\partial v} \nonumber\\
&=A\sqrt{\frac{1}{2\gamma \omega_1^2\omega_2}}(a_x+a_x^\dagger)+BC\sqrt{\frac{\gamma \omega_1}{2}}(a_v-a_v^\dagger) 
\end{align}
Hence, from Eq. \ref{eqxemq}
\begin{align}
e^{-Q/2}xe^{Q/2}\ket{0,0} &=A\sqrt{\frac{1}{2\gamma \omega_1^2\omega_2}}\ket{0,1}-BC\sqrt{\frac{\gamma \omega_1}{2}}\ket{1,0}\nonumber\\
 \bra{0,0}e^{-Q/2}xe^{Q/2}&=A\sqrt{\frac{1}{2\gamma \omega_1^2\omega_2}}\bra{0,1}+BC\sqrt{\frac{\gamma \omega_1}{2}}\bra{1,0}
\end{align}
Eq. \ref{propmatele} yields
\begin{align*}
G(\tau)&=A^2\frac{1}{2\gamma \omega_1^2\omega_2}\bra{0,1}e^{-\tau (H_0-E_0)}\ket{0,1}-(BC)^2\frac{\gamma \omega_1}{2}\bra{1,0}e^{-\tau (H_0-E_0)}\ket{1,0}\\
   &=\frac{1}{2\gamma \omega_1^2\omega_2}A^2e^{-\omega_2\tau }-\frac{\gamma \omega_1}{2}(BC)^2e^{-\omega_1\tau }
\end{align*}
Note that all the operators and state functions in equation above are defined solely in state space $\mathcal{V}_0$. However, the coefficients of the various matrix elements, in particular the negative sign on the second matrix element, are a result of the properties of state space $\mathcal{V}$ and could not have been generated by working solely in Hilbert space $\mathcal{V}_0$.

Eq. \ref{coeffab} yields
\begin{align*}
A^2&=\frac{\omega_1^2}{\omega_1^2-\omega_2^2}\\
(BC)^2&=\frac{1}{\gamma^2\omega_1^2\omega_2^2}\cdot\frac{\omega_2^2}{\omega_1^2-\omega_2^2}=\frac{1}{\gamma^2\omega_1^2}\cdot\frac{1}{\omega_1^2-\omega_2^2}
\end{align*}
Hence, collecting all the results yields the expected result that
\begin{align}
\label{propgeneral}
G(\tau)&=\frac{1}{2\gamma}\frac{1}{\omega_1^2-\omega_2^2}\left[\frac{e^{-\omega_2\tau}}{\omega_2}-\frac{e^{-\omega_1\tau}}{\omega_1}\right]
\end{align}

\section{Propagator: state space for $\omega_1 > \omega_2$} \label{propstatesp}
Recall from Eq. \ref{proeqnvev2}, that the propagator is given by
\begin{align}
G(\tau)&=\bra{\Psi_{00}}e^{-Q}xe^{-\tau (H-E_0)}x\ket{\Psi_{00}}\nonumber\\
\label{proeqnvevstsp}
&=\bra{\Psi^D_{00}}xe^{-\tau (H-E_0)}x\ket{\Psi_{00}}
\end{align}
The completeness equation for $H$, from Eq. \ref{completeeqn},  is given by
\begin{align}
%\label{completeeqn}
\mathbb{I}=\sum_{m,n=1}^\infty  \ket{\Psi_{mn}} \bra{\Psi^D_{mn}} 
\end{align}
and yields, from Eq. \ref{proeqnvevstsp}, the following
\begin{align}
   \label{propg1g21}
G(\tau)&=\sum_{m,n=1}^\infty\bra{\Psi^D_{00}}x  e^{-\tau (H-E_0)}\ket{\Psi_{mn}} \bra{\Psi^D_{mn}}x\ket{\Psi_{00}}\nonumber\\
   &=e^{-\tau \omega_1}\bra{\Psi^D_{00}}x \ket{\Psi_{10}} \bra{\Psi^D_{10}}x\ket{\Psi_{00}}+e^{-\tau \omega_2}\bra{\Psi^D_{00}}x \ket{\Psi_{01}} \bra{\Psi^D_{01}}x\ket{\Psi_{00}}\\
   \label{propg1g2}
   &=e^{-\tau \omega_1}G_1(\tau)+e^{-\tau \omega_2}G_2(\tau)
\end{align}

The vacuum state and its normalization, from Eq. \ref{vacqderivdual},  is the following
\begin{align*}
%\label{vacqderivdual} 
\Psi_{00}(x,v)&=N_{00}\psi_{00}(x,v)~~;~~\Psi^D_{00}(x,v)=\Psi_{00}(x,-v)=\Psi_{00}(-x,v)\\
\psi_{00}(x,v)&=\exp\{-\frac{\gamma}{2}(\omega_1+\omega_2)\omega_1\omega_2x^2-\frac{\gamma}{2}(\omega_1+\omega_2)v^2-\gamma\omega_1\omega_2xv\}\\
N_{00}&=(\omega_1\omega_2)^{1/4}\sqrt{\frac{\gamma}{\pi}(\omega_1+\omega_2)}
\end{align*}
Recall from Eq. \ref{firsttwostatessummary}, the following
\begin{align*}
%\label{firsttwostatessummary}
&\Psi_{10}(x,v) =N_{10}\left[v+ \omega_2 x \right]\psi_{00}(x,v)~;~
\Psi^D_{10}(x,v) =N_{10}\left[v- \omega_2 x \right]\psi^D_{00}(x,v) \nonumber\\
&\Psi_{01}(x,v) =N_{01}\left[v+ \omega_1 x \right]\psi_{00}(x,v)~;~
\Psi^D_{01}(x,v) =N_{01}\left[-v+ \omega_1 x \right]\psi^D_{00}(x,v)\nonumber\\
&N_{10}=\gamma\sqrt{2}\frac{(\omega_1+\omega_2)}{\sqrt{\pi(\omega_1-\omega_2)}}~\omega_1^{3/4}\omega_2^{1/4}~;~N_{01}=\gamma\sqrt{2}\frac{(\omega_1+\omega_2)}{\sqrt{\pi(\omega_1-\omega_2)}}~\omega_1^{1/4}\omega_2^{3/4}
\end{align*}
Using the coordinate representation for the state vectors yields the following.
\begin{align}
G_1(\tau)&=e^{-\tau \omega_1}\bra{\Psi^D_{00}}x \ket{\Psi_{10}} \bra{\Psi^D_{10}}x\ket{\Psi_{00}}\nonumber\\
     &=N^2_{10}N^2_{00}\int dxdv~ x(v+ \omega_2 x)\psi^D_{00}(x,v)\psi_{00}(x,v) \nonumber\\
     &~~~~~~~~~~~~~~~~~~~\times\int dxdv~x(v- \omega_2 x)\psi^D_{00}(x,v)\psi_{00}(x,v)\nonumber\\
     &=-N^2_{10}N^2_{00}\omega_2^2 \left[\frac{\pi}{2\gamma(\omega_1+\omega_2)^2(\omega_1\omega_2)^{3/2}}\right]^2 \nonumber\\
     \label{g1inter}
     &=-\frac{1}{2\gamma(\omega_1^2-\omega_2^2)^2\omega_1}     
\end{align}
Similarly
\begin{align}
G_2(\tau)&=e^{-\tau \omega_1}\bra{\Psi^D_{00}}x \ket{\Psi_{01}} \bra{\Psi^D_{01}}x\ket{\Psi_{00}}\nonumber\\
     &=N^2_{10}N^2_{00}\int dxdv~ x(v+ \omega_1 x)\psi^D_{00}(x,v)\psi_{00}(x,v) \nonumber\\
     &~~~~~~~~~~~~~~~~~~~\times\int dxdv~x(-v+ \omega_1 x)\psi^D_{00}(x,v)\psi_{00}(x,v)\nonumber\\
     &=N^2_{10}N^2_{00}\omega_1^2 \left[\frac{\pi}{2\gamma(\omega_1+\omega_2)^2(\omega_1\omega_2)^{3/2}}\right]^2 \nonumber\\
      \label{g2inter}
     &=\frac{1}{2\gamma(\omega_1^2-\omega_2^2)^2\omega_2}     
\end{align}
Hence, Eqs. \ref{propg1g2}, \ref{g1inter} \ref{g2inter} yield the expected result given in Eq. \ref{propgeneral}, namely that
\begin{align*}
%\label{propgeneral}
G(\tau)&=\frac{1}{2\gamma}\frac{1}{\omega_1^2-\omega_2^2}\left[\frac{e^{-\omega_2\tau}}{\omega_2}-\frac{e^{-\omega_1\tau}}{\omega_1}\right]
\end{align*}
There are a number of remarkable features of the state space derivation. The negative sign that appears in the propagator for the term $G_1(\tau)$ is usually taken to be a proof that no unitary theory can yield this result. The reason for this is the following; consider any arbitrary Hermitian Hamiltonian such that $H_A=H_A^\dagger$; the spectral resolution of this Hamiltonian in terms of its eigenstates $|\chi_{mn}\rangle$ is given by
\begin{align}
%\label{completeeqn}
\mathbb{I}=\sum_{mn=1}^\infty  \ket{\chi_{mn}} \bra{\chi_{mn}} 
\end{align}
Note that $\bra{\chi_{mn}}=\bra{\chi^D_{mn}}$ since for a Hermitian Hamiltonian $H_A$ the left and right eigenstate are complex conjugate of each other. Hence, the propagator for the Hermitian Hamiltonian $H_A$ is given by  
\begin{align}
G_A(\tau)&=\sum_{m,n=1}^\infty\bra{\chi^D_{00}}x  e^{-\tau (H_A-E_0)}\ket{\chi_{mn}} \bra{\chi^D_{mn}}x\ket{\chi_{00}}\nonumber\\
   &=e^{-\tau \omega_1}\bra{\chi^D_{00}}x \ket{\chi_{10}} \bra{\chi^D_{10}}x\ket{\chi_{00}}+e^{-\tau \omega_2}\bra{\chi^D_{00}}x \ket{\chi_{01}} \bra{\chi^D_{01}}x\ket{\chi_{00}}\nonumber\\
  % \label{propg1g2}
   &=e^{-\tau \omega_1}\big{|}\bra{\chi_{00}}x \ket{\chi_{10}} \big{|}^2+e^{-\tau \omega_2}\big{|}\bra{\chi_{00}}x \ket{\chi_{01}} \big{|}^2\nonumber
\end{align}
The result above shows that a Hermitian Hamiltonian defined on a Hilbert space cannot have  a propagator such as the one  given in Eq. \ref{propgeneral} except by allowing $\big{|}\bra{\chi_{00}}x \ket{\chi_{10}} \big{|}^2 <0$, which implies that $\ket{\chi_{10}}$ is a ghost state that has a negative norm. In contrast, the pseudo-Hermitian Euclidean Hamiltonian $H$ has a positive norm for all the states in its state space; the duality transformation in going from $\ket{\Psi_{mn}}$ to $\bra{\Psi^D_{mn}}$ provides the negative signs that allows for the propagator given in Eq. \ref{propgeneral}.

\section{Hamiltonian: Equal frequency limit} \label{sec:equalfreq}

In the equal frequency limit of $\omega_1=\omega_2$ the parameters of the $Q$-operator given in Eq. \ref{abqparamaet} become divergent and a well defined $Q$-operator  no longer exists. Although the Hamiltonian has special properties for the equal-frequency point, as discussed in Paper I the path integral is well behaved for all values of $\omega_1,\omega_2$, including the equal-frequency point at $\omega_1=\omega_2$. Moreover, the non-Hermitian Hamiltonian $H$ is also well defined in the equal frequency limit. 

The singularity for $Q$-operator is due to the fact that the acceleration Hamiltonian $H$ \textit{cannot} be mapped to an equivalent Hermitian Hamiltonian $H_0$. For the case of $\omega_1=\omega_2$,  non-Hermitian Hamiltonian $H$ is no longer pseudo-Hermitian but instead, $H$ is essentially non-Hermitian and has been shown to be expressible as a Jordan-block matrix  \cite{mannheim1}. 

The general analysis of the equal frequency Hamiltonian has been carried out for Minkowski time in the pioneering work of Bender and Mannheim \cite{mannheim1} and the analysis for Euclidean time is similar to their analysis, but with many details that are different. 

\section{Propagator and states for equal frequency}
To illustrate the general features of the equal frequency limit, the propagator is analyzed from the point of view of the underlying state space. As mentioned at the end of Section \ref{excitedstham},  in the limit of $\omega_1=\omega_2$ the single excitation eigenstates $\Psi_{10}, \Psi_{01}$ become degenerate, with both eigenstates having energy $2\omega$. 

The purpose of analyzing the propagator is to extract the state vectors that emerge in the limit of $\omega_1\to\omega_2$.

Since $\omega_1>\omega_2$, consider the limit of $\epsilon \to 0+$ with
\begin{align}
\label{pdefepsiomega}
\omega_1=\omega+\epsilon~~;~~\omega_2=\omega_2-\epsilon
\end{align}
which yields, from Eq. \ref{energymn}
\begin{align*}
E_{00}=\frac{1}{2}(\omega_1+\omega_2)\to\omega~~;~~E_{10}\to2\omega +\epsilon ~~;~~
E_{01}\to2\omega -\epsilon
\end{align*}

Consider the limit of $\omega_1 \to \omega_2$  for the state vector expression of the propagator given by Eq. \ref{propg1g21}
\begin{align}
\label{gtaubasci}
G(\tau)&=e^{-\tau \omega_1}\bra{\Psi^D_{00}}x \ket{\Psi_{10}} \bra{\Psi^D_{10}}x\ket{\Psi_{00}}+e^{-\tau \omega_2}\bra{\Psi^D_{00}}x \ket{\Psi_{01}} \bra{\Psi^D_{01}}x\ket{\Psi_{00}}\nonumber\\
   &=e^{-\tau \omega}[G_{10}+G_{01}]
\end{align}
where, defining $\int dxdvdx'dv'=\int_{x,v,x',v'}$ yields
\begin{align}
&G_{10}=e^{-\epsilon \tau}\bra{\Psi^D_{00}}x \ket{\Psi_{10}} \bra{\Psi^D_{10}}x\ket{\Psi_{00}}\nonumber\\
     &=N^2_{00}N^2_{10}\int_{x,v,x',v'}~ x(v+ \omega_2 x)\psi^D_{00}(x,v)\psi_{00}(x,v) ~x'(v'- \omega_2 x')\psi^D_{00}(x',v')\psi_{00}(x',v')\nonumber\\
     &=N^2_{00}N^2_{10}\int_{x,v,x',v'}~ xx'(v+ \omega_2 x) (v'- \omega_2 x')P(x,v)P(x',v')
\end{align}
and
\begin{align}
&G_{01}=e^{\epsilon \tau}\bra{\Psi^D_{00}}x \ket{\Psi_{01}} \bra{\Psi^D_{01}}x\ket{\Psi_{00}}\nonumber\\
     &=N^2_{00}N^2_{01}\int_{x,v,x',v'}~ x(v+ \omega_1 x)\psi^D_{00}(x,v)\psi_{00}(x,v) ~x'(-v'+ \omega_1 x')\psi^D_{00}(x',v')\psi_{00}(x',v')\nonumber\\
     &=N^2_{00}N^2_{01}\int_{x,v,x',v'}~ xx'(v+ \omega_1 x)~(-v'+ \omega_1 x')P(x,v)P(x',v')
\end{align}
where
\begin{align}
P(x,v)=\hat{\psi}^D_{00}(x,v)\hat{\psi}_{00}(x,v)=\exp\{-2\gamma\omega^3x^2-2\gamma\omega v^2\} \nonumber
\end{align}
since, in the limit of $\omega_1 \to \omega_2$ the vacuum state has the well defined limit given by
\begin{align}
\label{vaceqaul}
\lim_{\epsilon\to 0}\psi_{00}(x,v)=\hat{\psi}_{00}(x,v)=\exp\{-\gamma\omega^3x^2-\gamma\omega v^2-\gamma\omega^2 xv\} 
\end{align}

To leading order in $\epsilon$ the normalization constants yield the following
\begin{align}
\label{coeffc}
N^2_{10}\to C\frac{\omega_1}{\epsilon}~~;~~N^2_{10}\to C\frac{\omega_2}{\epsilon}~~;~~C=\frac{4\gamma^2\omega^3}{\pi} 
\end{align}
and from Eq. \ref{vacqderivnorm}
\begin{align}
\label{normvaceqaul}
\hat{N}^2_{00}=\lim_{\epsilon\to 0}N^2_{00}=\frac{2\gamma\omega^2}{\pi}
\end{align}

Hence, collecting above equations
\begin{align}
\label{gaux12}
G_{10}+G_{01}=C\hat{N}^2_{00}\int_{x,v,x',v'}~ xx'\mathcal{F}(x,v;x',v')P(x,v)P(x',v')
\end{align}
where
\begin{align}
\label{timeindep}
&\mathcal{F}(x,v;x',v')=e^{-\epsilon \tau}\frac{\omega_1}{\epsilon}(v+ \omega_2 x) (v'- \omega_2 x')+e^{\epsilon \tau}\frac{\omega_2}{\epsilon}(v+ \omega_1 x) (-v'+ \omega_1 x')
\end{align}
Expanding $\mathcal{F}(x,v;x',v')$ to leading order in $\epsilon$ yields the following
\begin{align}
\mathcal{F}(x,v;x',v')=\frac{1}{\epsilon}\Big[(1-\epsilon \tau)(\omega+\epsilon)\{v+ (\omega-\epsilon) x\} \{v'- (\omega-\epsilon) x'\}\nonumber\\
~~~~~~~~~~+(1+\epsilon \tau)(\omega-\epsilon)\{v+ (\omega+\epsilon) x\} \{-v'+ (\omega+\epsilon) x'\}\Big]+O(\epsilon)\nonumber\\
=2\Big[(\omega\tau-1)(v+ \omega x)(-v'+ \omega x')+ \omega x (-v'+ \omega x')+ \omega(v+ \omega x)x'\Big]+O(\epsilon)\nonumber\\
\label{fstatevecall}
=2\Big[-(v+ \omega x)(-v'+ \omega x')+ \omega \{x +\tau(v+ \omega x)\}(-v'+ \omega x')+ \omega(v+ \omega x)x'\Big]
\end{align}
The expression for $\mathcal{F}(x,v;x',v')$ given in Eq. \ref{fstatevecall} carries information of the state vectors that determine the propagator; to extract this information, the state vectors need to read off from above equation. Recall the state vectors and their duals are polynomials of $x,v$ multiplied into the vacuum state. Hence,  Eq. \ref{fstatevecall} yields the following \cite{mannheim1}
\begin{align}
\label{timeindep2}
&\psi_1(x,v;\tau)=\bra{x,v}\psi_1(\tau)\rangle=(v+ \omega x)\hat{\psi}_{00}(x,v)e^{-2\omega \tau}\\
\label{timeindep3}
&\psi^D_1(x,v;\tau)=\bra{\psi^D_1(\tau)}x,v \rangle=(-v+ \omega x)\hat{\psi}^D_{00}(x,v)e^{-2\omega \tau}\\
\label{timeindep4}
&\psi_2(x,v;\tau)=\bra{x,v}\psi_2(\tau)\rangle=\omega\{x +\tau(v+ \omega x)\}\hat{\psi}_{00}(x,v)e^{-2\omega \tau}\\
\label{timeindep5}
&\psi^D_2(x,v;\tau)=\bra{\psi^D_2(\tau)} x,v\rangle=\omega\{x +\tau(-v+ \omega x)\}\hat{\psi}^D_{00}(x,v)e^{-2\omega \tau}
\end{align}

The dual state vector is defined by $v\to -v$; namely\footnote{Defining the dual state vector by $x\to-x$, as is the case for Minkowski time \cite{mannheim1} gives an incorrect result for Euclidean time.}
\begin{align*}
&\psi^D_1(x,v;\tau)=\psi_1(x,-v;\tau)~~;~~\psi^D_2(x,v;\tau)=\psi_2(x,-v;\tau)
\end{align*}
Note the subtlety of conjugation for the unequal frequency case, with a different rule for each excited state as given in Eq. \ref{firsttwostatdual}, has been lost since the two excited states have become degenerate for the equal frequency case.

Collecting the results from Eqs. \ref{gtaubasci}, \ref{gaux12}, \ref{timeindep} - \ref{timeindep5} yields the following
\begin{align}
\label{gtautimedep}
G(\tau)&=2C\hat{N}^2_{00}e^{\tau \omega}\bra{\hat{\psi}^D_{00}}x \Big [-\ket{\psi_1(\tau)}\bra{\psi^D_1(0)}\nonumber\\
&~~~~~~~~~~~~~~~~~~~~~+\Big\{\ket{\psi_2(\tau)}\bra{\psi^D_1(0)}+\ket{\psi_1(\tau)}\bra{\psi^D_2(0)}\Big\} \Big]x\ket{\hat{\psi}_{00}}
\end{align}
The result in Eq. \ref{gtautimedep} shows that the eigenstates $\ket{\Psi_{10}},\ket{\Psi_{01}}$ that gave the result for the propagator in Eq. \ref{gtaubasci} have been replaced, in the limit of $\omega_1\to\omega_2$, by new state vectors that are well defined and finite for $\epsilon=0$.

In state vector notation Eqs. \ref{gtaubasci} and \ref{gtautimedep} yield the following 
\begin{align}
\label{gtaubascicompl}
&\lim_{\omega_1\to\omega_2}~~e^{-\tau E_{10}} \ket{\Psi_{10}} \bra{\Psi^D_{10}}+e^{-\tau E_{01}}\ket{\Psi^D_{00}} \bra{\Psi^D_{01}}\nonumber\\
   &=2C\hat{N}^2_{00} \Big [-\ket{\psi_1(\tau)}\bra{\psi^D_1(0)}+\ket{\psi_2(\tau)}\bra{\psi^D_1(0)}+\ket{\psi_1(\tau)}\bra{\psi^D_2(0)}\Big]
\end{align}

\section{Low energy state vectors for equal frequency} \label{sec:equalfreqstvec}
The Hamiltonian for the equal frequency case, from Eq. \ref{mainham}, is given by
\begin{equation}
\label{hamw1w2equal}
H=-\frac{1}{2\gamma}\frac{\partial^2}{\partial v^2}-v\frac{\partial}{\partial x}+\omega^2 v^2+\frac{\gamma}{2}\omega^4x^2
\end{equation}
The equal frequency vacuum state is an energy eigenstate with
\begin{equation}
%\label{hamw1w2}
H\hat{\psi}_{00}(x,v)=\omega\hat{\psi}_{00}(x,v)~~;~~\bra{\hat{\psi}^D_{00}} \hat{\psi}_{00}\rangle=\frac{1}{\hat{N}^{2}_{00}}=\frac{\pi}{2\gamma\omega^2}
\end{equation} 

The state vectors $\ket{\psi_1(\tau)}, \ket{\psi_2(\tau)}$ were obtained by analyzing the equal frequency propagator. The state vectors have the following interpretation.  

\subsection{State vector $\ket{\psi_1(\tau)}$}
The state vector $\ket{\psi_1(\tau)}$ is an energy eigenstate given by the average of the two unequal frequency eigenstates that become degenerate, namely
\begin{align}
\label{statevec1}
&\psi_1(x,v;\tau)=\lim_{\epsilon\to 0}\frac{1}{2}[e^{-\tau E_{10}}\psi_{10}(x,v)+ e^{-\tau E_{01}}\psi_{01}(x,v)]\nonumber\\
   &\Rightarrow \psi_1(x,v;\tau)\equiv\bra{x,v}\psi_1(\tau)\rangle=e^{-2\tau \omega}(v+\omega x)\hat{\psi}_{00}(x,v) \nonumber\\
 &H\psi_1(x,v;\tau)=2\omega\psi_1(x,v;\tau) 
\end{align}
The first sign of the irreducible non-Hermitian nature of the equal frequency Hamiltonian appears with $\ket{\psi_1(\tau)}$; unlike the norm of all the energy eigenstates, the norm of $\ket{\psi_1(\tau)}$ is zero; namely
\begin{equation}
\label{eigenzernorm}
\bra{\psi^D_1(\tau)} \psi_1(\tau)\rangle=0
\end{equation} 
The norm of the eigenstate being zero is a general feature of a Hamiltonian that is of the form of a Jordan-block and, in particular is not pseudo-Hermitian \cite{mannheim1}. The fact that the eigenstate has zero norm does not mitigate against the eigenstate being included in the collection of state vectors that, taken together, yield a resolution of the identity operator.

\begin{figure}[h]
  \centering
  \epsfig{file=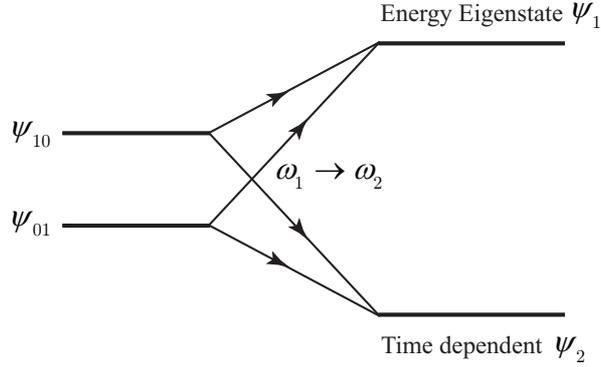, height=5cm}%, width=6cm}%, angle=0}
  \caption{The equal frequency limit yields two new states from two energy eigenstates $\Psi_{10}, ~\Psi_{10}$ of the un-equal frequency case.}
  \label{timedpstate}
\end{figure}

\begin{figure}
        \begin{subfigure}[b]{0.5\textwidth}
                \centering
                \vspace{-0.5cm}
                \includegraphics[width=\textwidth]{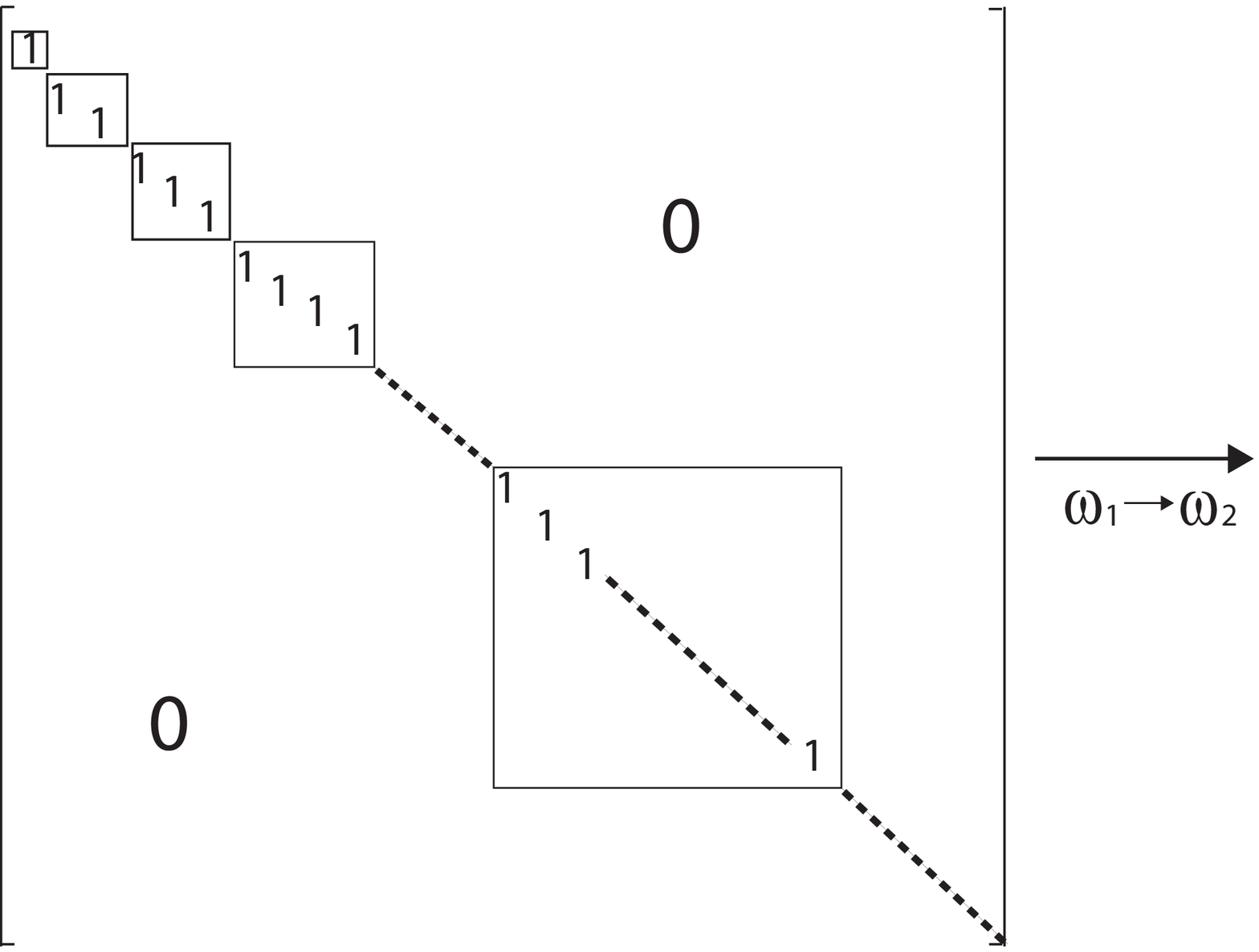}
                \caption{Unequal frequencies.}
                \label{fig:gull}
        \end{subfigure}%
        \hspace{0.05cm}  %add desired spacing between images, e. g. ~, \quad, \qquad etc. 
        \begin{subfigure}[b]{0.42\textwidth}
                \centering
                \includegraphics[width=\textwidth]{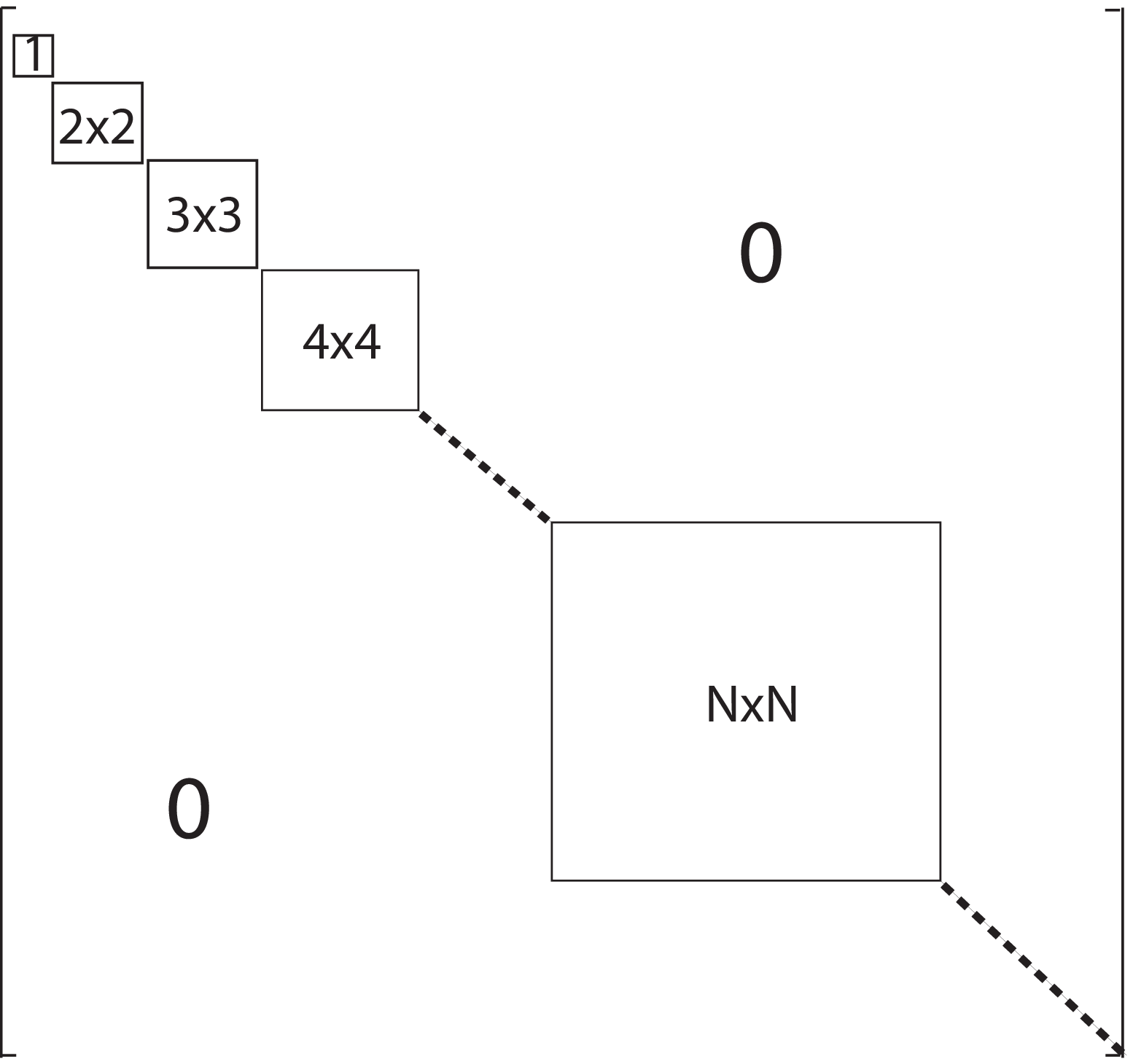}
                \caption{Equal frequencies.}
                %\label{fig:tiger}
        \end{subfigure}
       \caption{a) Completely diagonal Hamiltonian $H$ for the unequal frequency case. b) Block-diagonal structure of the Hamiltonian in the equal frequency limit, with each $N\times N$ block being given by a Jordan block.}
        \label{comeqn}
\end{figure}

\subsection{State vector $\ket{\psi_2(\tau)}$}
The second state vector $\ket{\psi_2(\tau)}$ that appears for the equal frequency case can be written as the difference of the two unequal frequency eigenstates that become degenerate; for dimensional consistency, the pre-factor of $\omega$ is introduced in the $\epsilon\to 0$; hence
\begin{align}
&\psi_2(x,v;\tau)=\lim_{\epsilon\to 0}\frac{\omega}{2\epsilon}\big[e^{-\tau E_{01}}\psi_{01}(x,v)-e^{-\tau E_{10}} \psi_{10}(x,v)\big]\nonumber\\
   &=\frac{\omega}{2\epsilon}e^{-2\omega\tau }\big[(1-\epsilon\tau)(v+(\omega-\epsilon) x)-(1-\epsilon\tau)(v+(\omega-\epsilon) x)\big]\hat{\psi}_{00}(x,v) \nonumber\\
\label{statevec2}   
&\Rightarrow \psi_2(x,v;\tau)=\bra{x,v}\psi_2(\tau)\rangle =e^{-2\tau \omega}\omega\big[x +\tau(v+ \omega x)\big]\hat{\psi}_{00}(x,v)
\end{align} 
Time-dependent state vector $\ket{\psi_2(\tau)}$ is not an (energy) eigenstate of $H$; however, since it results from the superposition of two energy eigenstates, it can be explicitly verified that $\ket{\psi_2(\tau)}$ satisfies the time dependent Schr\"odinger equation, namely
\begin{align}
\label{equalHaction}
&-\frac{\partial \psi_2(x,v;\tau)}{\partial \tau}=H\psi_2(x,v;\tau)~~\Rightarrow~~ \psi_2(x,v;\tau)=\exp\{-\tau H\}\psi_2(x,v;0)\nonumber\\
&\text{Initial value}~~~~~~\psi_2(x,v;0)= \omega x \hat{\psi}_{00}(x,v)
\end{align}
Note that $\ket{\psi_2(\tau)}$ has a finite norm and a non-zero overlap with the $\ket{\psi_1(\tau)}$; namely, using Eq. \ref{coeffc}
\begin{align}
\label{overlaptwo}
&\bra{\psi^D_2(\tau)}\psi_2(\tau)\rangle= \frac{e^{-4\tau \omega}}{2C}=\bra{\psi^D_2(\tau)}\psi_1(\tau)\rangle~~;~~C=\frac{4\gamma^2\omega^3}{\pi}
\end{align}

The equal frequency state space has a zero norm state, as in Eq. \ref{eigenzernorm}, and the time dependent state has a positive norm, unlike the case for Minkowski time  \cite{mannheim1} for which some of the state vectors have negative norm; in particular the norm of the time dependent state $|\psi_2(\tau)\rangle$ is positive definite. Of course, since one is working in Euclidean time probability is not conserved and one can see from Eq. \ref{overlaptwo} that the norm of the states decay exponentially to zero.

In summary, on taking the equal frequency limit, the two energy eigenstates $\ket{\Psi_{10}},\ket{\Psi_{01}}$ coalesce to yield a single energy eigenstate $\ket{\psi_1(\tau)}$; a second state time dependent state $\ket{\psi_2(\tau)}$ appears in this limit and takes the place of the loss of one of the eigenstates. The Hamiltonian is $2\times 2$ block diagonal matrix, as shown in Figure \ref{comeqn}. 

An analysis similar to the carried out for the single excitation level holds for all levels \cite{mannheim1}. The energy of the state $\ket{\Psi_{mn}}$, given in Eq. \ref{energymn}, has the following limit
\begin{align}
\label{equalenergies}
&E_{mn}=E_0+(m\omega_1+n\omega_2)\to \omega+(m+n)\omega~~;~~m,n=0,1,2,...\nonumber\\
&\Rightarrow E_N=N\omega~~;~~N=1,2,...
\end{align}
There are $N=1,2,3,..$ number of energy eigenstates at each level that  all collapse into a single (zero norm) energy eigenstate of the equal frequency Hamiltonian;  the single energy eigenstate has an energy equal to $E_N==N\omega$. 

In summary, the un-equal frequency Hamiltonian is completely diagonal, as shown in Figure \ref{comeqn}(a), and equivalent to a Hermitian Hamiltonian. When the equal frequency limit taken, the Hamiltonian is equal to an infinite dimensional block diagonal matrix, as shown in Figure \ref{comeqn}(b), with each block being composed of a $N\times N$ Jordan block matrices and is no longer a pseudo-Hermitian Hamiltonian.

All the $N$ eigenstates of the un-equal energy eigenstates collapse into a single eigenstate.  The $N-1$ eigenstates that are `lost' are replaced by $N-1$ time-dependent state vectors that are the superposition of the eigenstates of the unequal frequency Hamiltonian. For energy level $N\omega$, the time-dependent states together with the single eigenstate provide a resolution of the identity. This structure of the equal frequency Hamiltonian operator is illustrated in Figure \ref{comeqn}(b). 

\section{Completeness equation for $2\times 2$ block}\label{sec:completenesequal}
We now discuss how the time-dependent state replaces the lost energy eigenstate to provide the complete set of states for the equal frequency case.

The example of the \textit{single excitation} states, created by applying a creation operator $a_v^\dagger$ or $a_x^\dagger$ to the harmonic oscillator vacuum state $\ket{0,0}$, showed that in the limit of $\omega_1=\omega_2$ the two energy eigenstates $\ket{\Psi_{10}},\ket{\Psi_{01}}$ were superposed to create new states $\ket{\psi_1(\tau)},\ket{\psi_2(\tau)}$. 

Since the orthogonality of the eigenstates is maintained in the superposition the mixing of states is only amongst states of a fixed excitation; in other words, states having \textit{two excitations} consisting of applying the creation operator two times, namely $(a_v^\dagger)^2$, $(a_x^\dagger)^2$ or $a_v^\dagger a_x^\dagger$ yield three eigenstates that only mix with each other in the limit of $\omega_1=\omega_2$. And so on for all the higher excitations states.

Hence, the resolution of the identity -- which is an expression of the completeness of a set of basis states --  as shown in Figure \ref{comeqn}, breaks up into a block-diagonal form, with states of a given excitation mixing with each other and not with the states of the other blocks.

To illustrate the general result, consider the $2\times 2$ block for the single excitation states. In light of the result obtained in Eq. \ref{gtautimedep}, consider the following Hermitian anstaz for the  $2\times 2$ block identity operator, with all the state vectors taken at initial time $\tau=0$. For notational simplicity, let 
\begin{align}
\label{psipsi2equal1}
&\ket{\psi_1(0)}=\ket{\psi_1}=(v+\omega x)\ket{\hat{\psi}_{00}}~~:~~\bra{\psi^D_1(0)}=\bra{\psi^D_1}\\
\label{psipsi2equal2}
&\ket{\psi_2(0)}=\ket{\psi_2}=x\ket{\hat{\psi}_{00}}~~;~~ \bra{\psi^D_2(0)}=\bra{\psi^D_2}
 \end{align}
Then, the identity operator, which is Hermitian, has the following representation for the 2$\times$2 block of Hilbert space
\begin{align}
\label{twobytwoiden}
&\mathbb{I}_{2\times 2}=-P\ket{\psi_1}\bra{\psi^D_1}+Q\Big[\ket{\psi_2}\bra{\psi^D_1}+\ket{\psi_1}\bra{\psi^D_2}\Big]
\end{align}
Recall from Eqs. \ref{eigenzernorm} and \ref{overlaptwo}
\begin{align*}
%\label{overlaptwo}
&\bra{\psi^D_1} \psi_1\rangle=0~~;~~\bra{\psi^D_2}\psi_2\rangle= \frac{1}{2C}=\bra{\psi^D_2}\psi_1\rangle
\end{align*}
Eq. \ref{twobytwoiden} requies that
\begin{align*}
%\label{twobytwoiden}
\mathbb{I}_{2\times 2}&=\mathbb{I}^2_{2\times 2}\nonumber\\
&=\frac{1}{2C}\left\{\big(-2PQ+Q^2\big)\ket{\psi_1}\bra{\psi^D_1}+Q^2\Big[\ket{\psi_2}\bra{\psi^D_1}+\ket{\psi_1}\bra{\psi^D_2}\Big]\right\}
\end{align*}
and yields from Eqs. \ref{twobytwoiden} and \ref{overlaptwo}
\begin{align*}
%\label{overlaptwo}
&P=\frac{1}{2C}\big(2PQ-Q^2\big)~~;~~Q=\frac{1}{2C}Q^2\\
&\Rightarrow P=Q=2C= \frac{8\gamma^2\omega^3}{\pi}
\end{align*}
Hence, the completeness equation for the $2\times 2$ block single excitation states is given by
\begin{align}
\label{twobytwoidenfinal}
\mathbb{I}_{2\times 2} ~~~~~~~~~~~~~~~~~~~~~~~~~~~~~~~~~~~\nonumber\\
=2C\Big[-\ket{\psi_1(0)}\bra{\psi^D_1(0)}+\ket{\psi_2(0)}\bra{\psi^D_1(0)}+\ket{\psi_1(0)}\bra{\psi^D_2(0)}\Big]
\end{align}
The completeness equation above is equal, up to a normalization, to Eq. \ref{gtaubascicompl}.
\section{Equal frequency propagator}
The defining equation for the propagator is, from Eq. \ref{proeqnvevstsp}, the following
\begin{align}
G(\tau)&=\bra{\Psi^D_{00}}xe^{-\tau (H-\omega)}x\ket{\Psi_{00}}\nonumber\\
\label{propequalfreq}
\Rightarrow \hat{G}(\tau)&=\lim_{\epsilon\to 0}G(\tau) =\hat{N}^2_{00}\bra{\hat{\psi}^D_{00}}xe^{-\tau (H-E_0)}x\ket{\hat{\psi}_{00}}
\end{align}

The completeness equation can be used to give a derivation  of the equal frequency propagator from first principles. Inserting the completeness equation given in Eq. \ref{twobytwoidenfinal} into the expression for the equal frequency propagator given in Eq. \ref{propequalfreq} yields the following
\begin{align}
\label{propghat}
\hat{G}(\tau)=2C\hat{N}^2_{00} \bra{\hat{\psi}^D_{00}}x
e^{-\tau (H-\omega)} ~~~~~~~~~~~~~~~~~~~~~~~~~~~ \nonumber\\
\times\Big[-\ket{\psi_1(0)}\bra{\psi^D_1(0)}+\ket{\psi_2(0)}\bra{\psi^D_1(0)}+\ket{\psi_1(0)}\bra{\psi^D_2(0)}\Big]x\ket{\hat{\psi}_{00}}
\end{align} 
Note Eq. \ref{propghat} above is equivalent to the earlier expression given in Eq. \ref{gtautimedep}.

It follows from Eqs. \ref{statevec1} and \ref{statevec2} that 
\begin{align}
\label{equalstavec1}
&\bra{x,v}e^{-\tau H}\ket{\psi_1(0)}=\langle x,v\ket{\psi_1(\tau)}=e^{-2\tau \omega}(v+\omega x)\hat{\psi}_{00}(x,v)\\
\label{equalstavec2}
&\bra{x,v}e^{-\tau H}\ket{\psi_2(0)}=\langle x,v\ket{\psi_2(\tau)}\nonumber\\
&~~~~~~~~~~~~~~~~~~~~~~= e^{-2\tau \omega}\omega\{x+\tau(v+\omega x)\}\hat{\psi}_{00}(x,v)
\end{align}

It can be shown from the first and last term inside the square bracket in Eq. \ref{propghat} cancel. Hence, from Eqs. \ref{propghat}, \ref{equalstavec1} and \ref{equalstavec2} 
\begin{align}
%\label{propghat}
&\hat{G}(\tau)=e^{-\tau \omega}2C\hat{N}^2_{00} \bra{\hat{\psi}^D_{00}}x\ket{\psi_2(\tau)}\bra{\psi^D_1(0)}x\ket{\hat{\psi}_{00}}\nonumber\\
    &=e^{-\tau \omega} 2C\hat{N}^2_{00}\int dxdv~ \omega x\{x+\tau(v+\omega x)\}P(x,v)\cdot\int dxdv~ x(-v+\omega x)P(x,v)\nonumber\\
    &=e^{-\tau \omega} 2C \omega^2\hat{N}^2_{00}  \left[1+\omega\tau\right]\left[\int dxdv~ x^2P(x,v)\right]^2 
    \end{align}
Performing the Gaussian integrations yields
\begin{align}
\label{xsqexpecval}
&\int dxdv~ x^2P(x,v) =\frac{1}{2\omega^2 C}
    \end{align}    
Hence    
    \begin{align}
 \label{equalfinalprop}
&\hat{G}(\tau) =e^{-\tau \omega}\frac{\hat{N}^2_{00}}{2\omega^2C} [1+\omega\tau] =\frac{1}{4\gamma}\frac{1}{\omega^3}e^{-\omega\tau}\left[1+\omega\tau\right]
\end{align}
where $C=(4\gamma^2\omega^3)/\pi$ is given in Eq. \ref{coeffc} and the normalization constant $\hat{N}^2_{00}=2\gamma\omega^2/\pi$ is given in Eq. \ref{normvaceqaul}.

\begin{figure} [h]
        \begin{subfigure}[b]{0.5\textwidth}
                \centering
                \includegraphics[width=\textwidth]{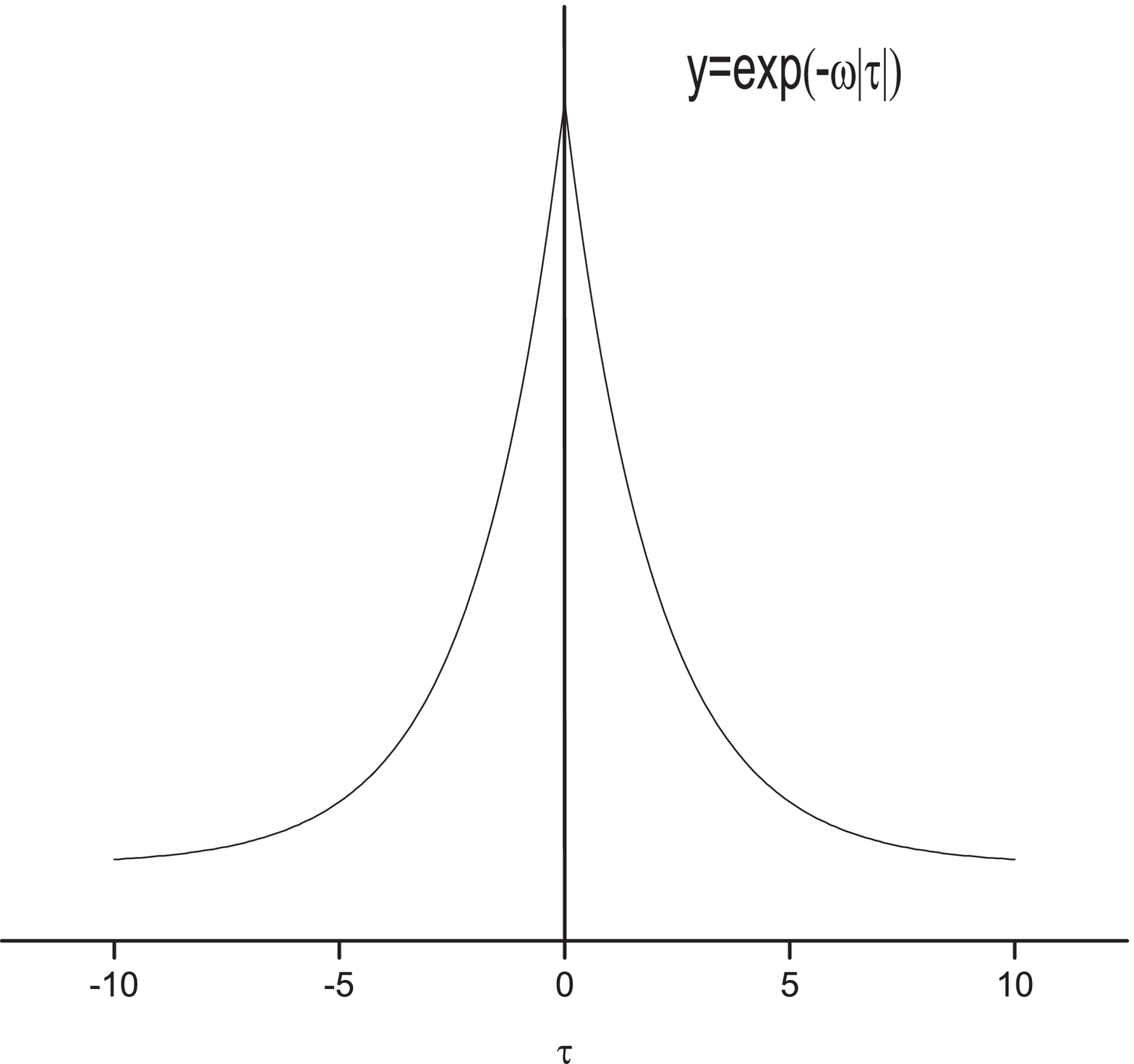}
                \caption{Single Exponential $\exp\{-\omega |\tau|\}$.}
                %\label{fig:gull}
        \end{subfigure}%
        \hspace{1.0cm}  %add desired spacing between images, e. g. ~, \quad, \qquad etc. 
        \begin{subfigure}[b]{0.5\textwidth}
                \centering
                \includegraphics[width=\textwidth]{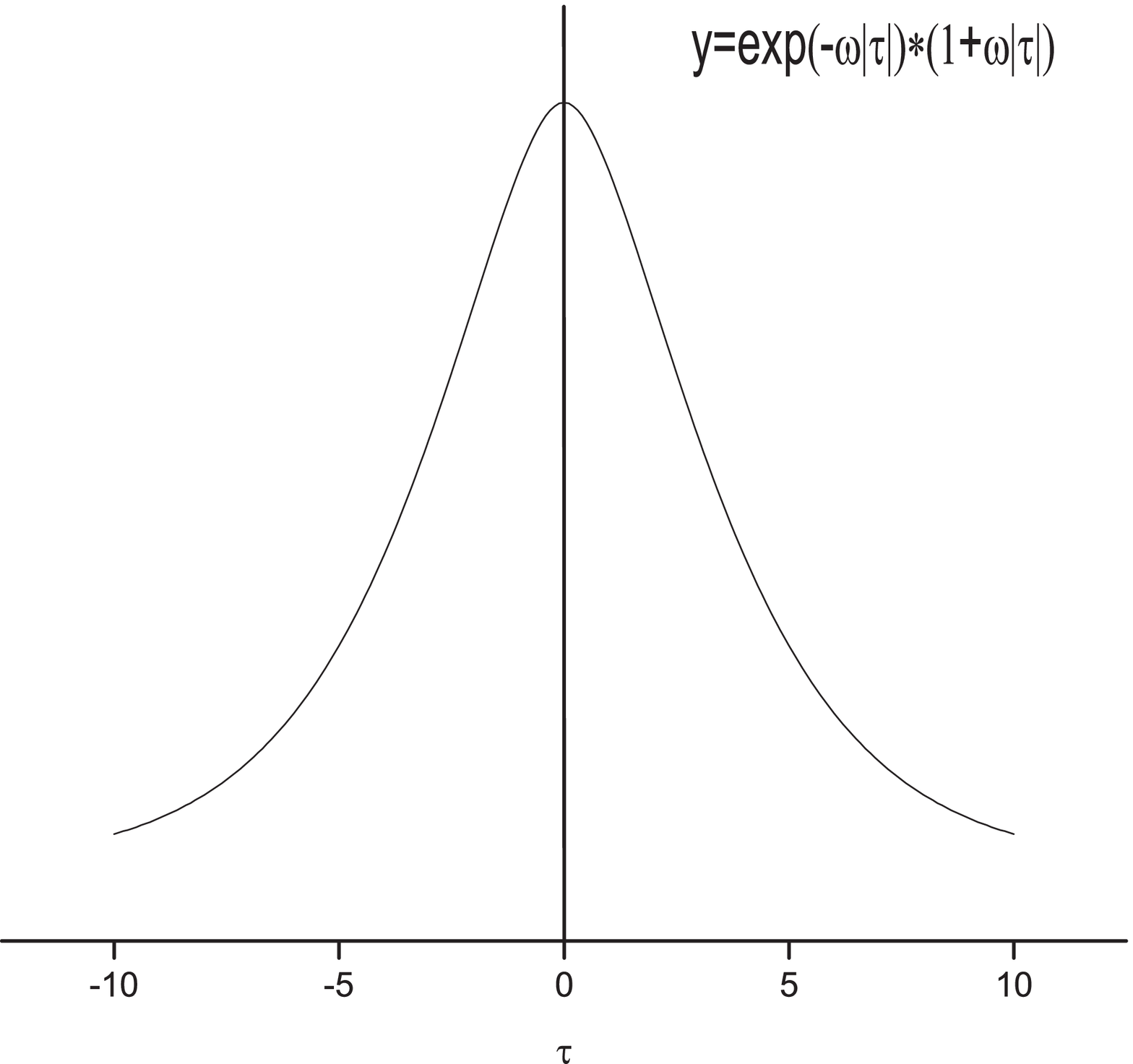}
                \vspace{-1.0cm}
                \caption{Propagator for equal frequency ~~~~~~~~~~~ $\exp\{-\omega |\tau|\}[1+\omega |\tau|]$.}
                \label{fig:tiger}
        \end{subfigure}
        %\caption{Pictures of animals}\label{fig:animals}
\end{figure}

To verify the equal frequency result obtained for the propagator, consider taking the limit of $\omega_1 \to \omega_2$ in Eq. \ref{propgeneral}. The propagator  has the following well-defined and finite limit
\begin{align}
\label{propsing}
\hat{G}(\tau)&=\lim_{\epsilon\to 0}\frac{1}{4\gamma\epsilon}\frac{1}{\omega_1+\omega_2}e^{-\omega\tau}
\left[\frac{e^{\epsilon\tau}}{\omega-\epsilon}-\frac{e^{-\epsilon\tau}}{\omega+\epsilon}\right] \nonumber\\
&=\frac{1}{4\gamma}\frac{1}{\omega^3}e^{-\omega\tau}\left[1+\omega\tau\right]
\end{align}
and agrees with the result obtained in Eq. \ref{equalfinalprop}.

Figures \ref{fig:gull} and \ref{fig:tiger} shows a comparison between the equal frequency propagator and the exponential function; the kink for the exponential function at $\tau=0$ is smoothed out for the equal frequency propagator.

\section{Jordan block structure} \label{sec:jordanblock}
In the limit of equal frequencies $\omega_1=\omega_2$ there is a re-organization of state space into a direct sum of finite dimensional subspaces, one subspace for each block diagonal component of $H$, as shown in Figure \ref{comeqn}. The break-down of the pseudo-Hermitian property of the Hamiltonian $H$ is due to the fact that, for equal frequencies, $H$ becomes a direct of sum of Jordan blocks. 

The total Hilbert space $\mathcal V$ breaks up into a direct sum of finite dimensional vector spaces $\mathcal V_n$ and is given by
\begin{align}
\mathcal V=\oplus_{n=1}^\infty \mathcal V_n
\end{align}
where $\mathcal V_1$ is one dimensional, $\mathcal V_2$ is two dimensional and so on.

The Hamiltonian is a direct sum of finite dimensional block matrices, denoted  by matrix $\mathcal H_n$, shown in Figure \ref{comeqn}, that acts on the subspace $\mathcal V_n$; the Hamiltonian is given by the following block diagonal decomposition
\begin{eqnarray}
\label{hjpordansum}
\mathcal H&=\oplus_{n=1}^\infty \mathcal H_n =\oplus_{n=1}^\infty a_n\mathcal J_{\lambda_n ,n}
\end{eqnarray}
The coefficients $a_n$ are real constants; $\mathcal J_{\lambda_n ,n}$ is a $n\times n$ Jordan block -- specified by its size $n$ and eigenvalue $\lambda_n $ -- and is given by\footnote{The $\pm 1$ terms in the super-diagonal in Eq. \ref{jordannn} are allowed since multiplying $\mathcal J_{\lambda_n ,n}$ by $-1$ can switch the sign the super-diagonal  from $1$  to and $-1$, and in doing so re-define the eigenvalue to be $-\lambda_n $.}
\begin{eqnarray}
\label{jordannn}
\mathcal J_{\lambda_n ,n}=\left[%
\begin{array}{cccccc}
  \lambda_n & \pm 1  & 0 & 0 & \ddots &\ddots\\
   0& \lambda_n & \pm 1 & 0 & \ddots &\ddots\\
   \ddots&  0& \lambda_n & \pm 1  & 0 &\ddots\\
   \ddots& \ddots & \ddots & \ddots & \ddots &\ddots\\
   \ddots& \ddots &\ddots  & 0 & \lambda_n & \pm 1\\
   \ddots& \ddots & \ddots & \ddots & 0  &   \lambda_n\\
\end{array}%
\right]
\end{eqnarray}

The Hamiltonian is analyzed for the first two blocks;  $\mathcal H_1$ is one dimensional and $\mathcal H_2$ is a 2$\times$2 matrix.

The ground state forms an invariant subspace   $\mathcal V_1$ with a single element $|e_0\rangle$ proportional to $|\hat{\psi}_{00}\rangle$; for dimensional consistency and to preserve the correct normalization, the following is the mapping
\begin{align}
\label{ezero}
|e_0\rangle=\hat{N}_{00}|\hat{\psi}_{00}\rangle~~;~~\langle e_0|e_0\rangle=1
\end{align}
The eigenvalue equation  $H|\hat{\psi}_{00}\rangle=\omega |\hat{\psi}_{00}\rangle$ yields the Hamiltonian on $\mathcal V_1$ given by
\begin{eqnarray}
\label{hamzero}
\mathcal H_1=\omega~~;~~\mathcal H_1|e_0\rangle=\omega |e_0\rangle
\end{eqnarray}

\section{2$\times$2 Jordan block}
A derivation is given using the 2$\times$2 Jordan block structure of the Hamiltonian and state space.

The result given in Eq. \ref{hamzero} together with Eq. \ref{hjpordansum} yields
\begin{align}
%\label{hjpordansum}
\mathcal H&=\mathcal H_1+\mathcal H_2+... \\
&=\omega \oplus 2\omega \mathcal J_{2}\oplus ..
          \end{align}
It will be shown in this Section that
\begin{eqnarray}
\label{jordan22}
\mathcal J_{2}=\left[%
\begin{array}{cc}
  1 &-1  \\
   0& 1 
   \end{array}%
\right]
\end{eqnarray}

Bender and Mannheim \cite{mannheim1} derive the 2$\times$2 Jordan block for the Minkowski Hamiltonian by defining creation and destruction operators that have a finite limit when $\epsilon\to 0$. In this Section, the 2$\times$2 Jordan block for the Euclidean Hamiltonian is directly derived from the state vectors and completeness obtained by taking the $\epsilon\to 0$, as discussed in Sections \ref{sec:equalfreqstvec} and \ref{sec:completenesequal}.

Recall from Eqs. \ref{hamw1w2equal}, \ref{psipsi2equal1} and \ref{psipsi2equal2}, the Hamiltonian and state vectors for the equal frequency limit are given by 
\begin{align*}
&H=-\frac{1}{2\gamma}\frac{\partial^2}{\partial v^2}-v\frac{\partial}{\partial x}+\omega^2 v^2+\frac{\gamma}{2}\omega^4x^2\\
&\ket{\psi_1}=\ket{\psi_1(0)}=(v+\omega x)\ket{\hat{\psi}_{00}}~~:~~\bra{\psi^D_1}=\bra{\psi^D_1(0)}\\
&\ket{\psi_2}=\ket{\psi_2(0)}=\omega x\ket{\hat{\psi}_{00}}~~;~~ \bra{\psi^D_2}=\bra{\psi^D_2(0)}
\end{align*}
The fact that the state vectors $\ket{\psi_1},~\ket{\psi_2}$ form a closed subspace under the action of $H$ points to an invariant 2$\times$2 subspace of the total Hilbert space.

In the 2$\times$2  block space, the Hamiltonian can be represented by a 2$\times$2 Jordan block in a basis fixed by the representation of $\ket{\psi_1},\ket{\psi_2}$ by 2 dimensional column vectors. To obtain this finite dimensional representation, note that $H\ket{\hat{\psi}_{00}}=\omega \ket{\hat{\psi}_{00}}$ and from Eq. \ref{statevec1} $H\ket{\psi_{1}}=2\omega \ket{\psi_{1}}$; hence, the action of $H$ on the state vectors $\ket{\psi_1},\ket{\psi_2}$ is given as follows 
\begin{align}
\label{firstequaleqn}
&H\ket{\psi_1}=2\omega \ket{\psi_1}\\
&H\ket{\psi_2}=-\omega v\ket{\hat{\psi}_{00}}+\omega^2 x \ket{\hat{\psi}_{00}}=-\omega(v+\omega x)\ket{\hat{\psi}_{00}}+2\omega  \ket{\psi_2}\nonumber\\
\label{secondequaleqn}
&\Rightarrow H\ket{\psi_2}=-\omega\ket{\psi_1}+2\omega  \ket{\psi_2}
\end{align}

Since $\ket{\psi_1}$ is an eigenvector of the Jordan block it is natural to make the following identification 
\begin{eqnarray}
\ket{\psi_1}\propto \left[%
\begin{array}{c}
  1   \\
   0 
   \end{array}%
\right]
\end{eqnarray}
Recall from Eqs. \ref{eigenzernorm} and \ref{overlaptwo}
\begin{align}
\label{overlaptwoe1e2}
&\bra{\psi^D_1} \psi_1\rangle=0~~;~~\bra{\psi^D_2}\psi_2\rangle= \frac{1}{2C}=\bra{\psi^D_2}\psi_1\rangle
\end{align}
Since $|\psi_1\rangle$ has zero norm, its normalization is fixed by its overlap with $|\psi_2\rangle$. Choosing the normalization consistent with above Eq. \ref{overlaptwoe1e2} yields the following 
\begin{eqnarray}
\label{eiggenequal}
\sqrt{C}\ket{\psi_1}=\ket{e_1}= \left[%
\begin{array}{c}
  1   \\
   0 
   \end{array}%
\right]~~;~~\sqrt{C}\ket{\psi_2}=\ket{e_2}= \left[%
\begin{array}{c}
  1/2   \\
   1/2 
   \end{array}%
\right]
\end{eqnarray}
with the dual vectors given by
\begin{eqnarray}
\label{eiggenequaldual}
\sqrt{C}\bra{\psi^D_1}=\bra{e^D_1}= \left [%
\begin{array}{c}
  0 ~,~
   1 
   \end{array}\right]~~;~~\sqrt{C}\bra{\psi^D_2}=\bra{e^D_2}= \left[%
\begin{array}{c}
  1/2   ~,~
   1/2 
   \end{array} \right]
\end{eqnarray}
Note $\bra{e^D_1}$ is \textit{not} the transpose of $\ket{e_1}$.

The completeness equation for the state space of the 2$\times$2 block has a discrete realization; recall from Eq. \ref{twobytwoidenfinal} 
\begin{align}
\label{2dimcomple}
\mathbb{I}_{2\times 2}&=2C\Big[-\ket{\psi_1}\bra{\psi^D_1}+\ket{\psi_2}\bra{\psi^D_1}+\ket{\psi_1}\bra{\psi^D_2}\Big] \nonumber\\
\Rightarrow\mathbb{I}_{2\times 2}&=2\Big[-\ket{e_1}\bra{e^D_1}+\ket{e_2}\bra{e^D_1}+\ket{e_1}\bra{e^D_2}\Big]
\end{align}
The completeness equation for the Jordan block shows that there is an effective metric on the discrete state space $\mathcal V_2$. 

Using Eqs. \ref{eiggenequal} and \ref{eiggenequaldual}, Eq. \ref{2dimcomple} yields the following 
\begin{align*}
\mathbb{I}_{2\times 2}&=2\left\{-\left[%
\begin{array}{cc}
 0  &1  \\
   0& 0 
   \end{array}%
\right]+\frac{1}{2}\left[%
\begin{array}{cc}
  1 &1  \\
   0& 0 
   \end{array}%
\right]+\frac{1}{2}\left[%
\begin{array}{cc}
  0 &1  \\
   0& 1 
   \end{array}%
\right]\right\}=\left[%
\begin{array}{cc}
  1 &0  \\
   0& 1 
   \end{array}%
\right]
\end{align*}
and we have obtained the expected result. 

\subsection{Hamiltonian}
Let $\mathcal H_2$ denote the realization of the Hamiltonian as a discrete and dimensionless matrix action on the 2-dimensional state space of the 2$\times$2 Jordan block. Applying Eq. \ref{eiggenequal} to Eqs. \ref{firstequaleqn} and \ref{secondequaleqn} yields the following 2$\times$2 representation 
\begin{align*}
%\label{firstequaleqn}
&\mathcal H_2\ket{e_1}=2\omega \ket{e_1}~~\Rightarrow~~\bra{e^D_1}\mathcal H_2=2\omega \bra{e^D_1}\\
& \mathcal H_2\ket{e_2}=-\omega \ket{e_1}+2\omega  \ket{e_2}~~\Rightarrow~~\bra{e^D_2}\mathcal H_2=-\omega \bra{e^D_1}+2\omega  \bra{e^D_2}
\end{align*}
The Hamiltonian $\mathcal H_2$ --  in the $\ket{e_1}$ and $\ket{e_2}$ basis --  is proportional to the 2$\times$2 Jordan block matrix and is given by\footnote{The Euclidean Hamiltonian given in Eq. \ref{jordan22} has a -1 for the superdiagonal, unlike the case for the Minkowski Hamiltonian \cite{mannheim1} where it is +1.}
\begin{eqnarray}
%\label{jordan22}
\mathcal H_2=2\omega\left[%
\begin{array}{cc}
  1 &-1  \\
   0& 1 
   \end{array}%
\right]
\end{eqnarray}
The definition of the discrete vectors $ \ket{e_1}$  and $ \ket{e_1}$ given in Eq. \ref{eiggenequal} requires a rescaling by $\sqrt C$ due to dimensional consistency; in contrast, there is no need to rescale $\mathcal H_2$ since it has correct dimension  set by $\omega$.

The Jordan block Hamiltonian given in Eq. \ref{jordan22} has only one eigenvalue and this is the reason that the two different eigenstates for the unequal frequencies collapsed into a single eigenstate. The Jordan block limit of $\mathcal H_2$ (for equal frequency) shows that $\mathcal H_2$ is no longer pseudo-Hermitian since the Jordan block is inequivalent to any Hermitian matrix.

The right eigenvector of $\mathcal H_2$ is $\ket{e_1}$ and the left eigenvector of $\mathcal  H$ is the dual  $\bra{e^D_1}$; namely
\begin{align*}
&\mathcal H_2\ket{e_1}=2\omega \ket{e_1} ~~;~~\bra{e^D_1}\mathcal H_2= 2\omega \bra{e^D_1}\\
&\Rightarrow \bra{e^D_1} e_1\rangle=0=\bra{\psi^D_1} \psi_1\rangle
\end{align*}
Hence, the Jordan block structure shows why the equal frequency eigenstate has a zero norm. 

\subsection{Schrodinger equation  for Jordan block}
The Schrodinger equation for an arbitrary vector $\ket{e}$ is given 
\begin{align*}
%\label{firstequaleqn}
&-\frac{\partial}{\partial \tau}\ket{e(\tau)}= \mathcal H_2\ket{e(\tau)}
\end{align*}
For eigenvector $\ket{e_1}$ the time dependent solution is
\begin{align}
\label{firstequaleqneigen}
&-\frac{\partial}{\partial \tau}\ket{e_1(\tau)}= \mathcal H_2\ket{e_1(\tau)}=2\omega\ket{e_1(\tau)} \nonumber\\
&\Rightarrow \ket{e_1(\tau)}=e^{-2\omega\tau}\ket{e_1}~~;~~\ket{e_1}= \left[%
\begin{array}{c}
  1   \\
   0 
   \end{array}%
\right]
\end{align}  
The time-dependence of the state vector $\ket{e_2(\tau)}$ is given by the following
\begin{align}
\label{sche2}
&-\frac{\partial}{\partial \tau}\ket{e_2(\tau)}= \mathcal H_2\ket{e_2(\tau)}~~;~~\ket{e_2(0)}=\ket{e_2}= \left[%
\begin{array}{c}
  1/2   \\
   1/2 
   \end{array}%
\right]
\end{align} 
In the 2$\times 2$ block representation $\ket{e_2(\tau)}$ is given from the solution obtained in Eq. \ref{statevec2}, which yields 
\begin{align}
\psi_2(x,v;\tau) &=e^{-2\tau \omega}\omega \big[x +\tau(v+ \omega x)\big]\hat{\psi}_{00}(x,v) \nonumber\\
%\Rightarrow \sqrt{C} \ket{e_2(\tau)}&=e^{-2\tau \omega}\Big[\sqrt{C} \ket{e_2}+\tau \omega \sqrt{C}\ket{e_1}\Big] \nonumber\\
\label{sche2soln}
 \Rightarrow  \ket{e_2(\tau)}&=e^{-2\tau \omega}\Big[\ket{e_2}+\omega\tau \ket{e_1}\Big]   =e^{-2\tau \omega}\left[%
\begin{array}{c}
  1/2 +\omega \tau  \\
   1/2 
   \end{array}%
\right]
\end{align} 
It can be directly verified using the explicit form for the Hamiltonian given in Eq. \ref{jordan22} that the solution for $\ket{e_2(\tau)}$  given in Eq. \ref{sche2soln} satisfies the Schrodinger equation given in Eq. \ref{sche2}.

\subsection{Time evolution}
The Jordan block Hamiltonian is given by Eq. \ref{jordan22}; a simple calculation yields the evolution operator
\begin{eqnarray}
\label{jordan22time}
e^{-\tau \mathcal H_2}=  e^{-2\omega\tau}\left[%
\begin{array}{cc}
  1 &2\omega\tau  \\
   0& 1 
   \end{array}%
\right]
\end{eqnarray}

The time dependence of the state vectors follow directly from the evolution operator. For eigenvector $\ket{e_1}$ the time dependent solution is
\begin{align*}
%\label{firstequaleqn}
&\ket{e_1(\tau)}= e^{-\tau \mathcal H_2}\ket{e_1}=e^{-2\omega\tau}\ket{e_1}
\end{align*}  
which is the expected result as in Eq. \ref{firstequaleqneigen}.

The time-dependence of the state vector $\ket{e_2(\tau)}$ is given by the following
\begin{align}
\label{sche2}
&\ket{e_2(\tau)}= e^{-\tau \mathcal H_2}\ket{e_2} =e^{-2\omega\tau }\left[%
\begin{array}{c}
  1/2 +\omega \tau  \\
   1/2 
   \end{array}%
\right]
\end{align}
which is the expected result as in Eq. \ref{sche2soln}.

\section{Jordan Block Propagator}
The  equal frequency propagator is given in Eq. \ref{propequalfreq}
\begin{align*}
%\label{propghat}
&\hat{G}(\tau)=\hat{N}^2_{00} \bra{\hat{\psi}^D_{00}}X
e^{-\tau (H-\omega)}X\ket{\hat{\psi}_{00}}
\end{align*}

The position operator $X$, unlike the Hamiltonian, is not block diagonal for the equal frequency case; to determine the propagator, the representation of the position operator $X$ needs to determined in the 3$\times$3 subspace given by $\mathcal V_1\oplus\mathcal V_2$, which includes the ground state and the 2$\times 2$  Jordan block. The operator $X$ has the following matrix elements
\begin{align}
\label{xmatrixelements}
&\bra {\hat{\psi}_{00}^D}X\ket{\hat{\psi}_{00}}=0=\bra {\psi_{1}^D}X\ket{\psi_{1}}=\bra {\psi_{2}^D}X\ket{\psi_{2}}=\bra {\psi_{2}^D}X\ket{\psi_{1}} \nonumber\\
&\bra {\hat{\psi}_{00}^D}X\ket{\psi_{1}}=\frac{1}{2\omega C}=\bra {\hat{\psi}_{00}^D}X\ket{\psi_{2}}
\end{align}
Note the matrix elements of the operator $X$ are zero within a block and are non-zero only for elements that  connect vectors from two different blocks.

Since $X$ acts on the $\mathcal V_1\oplus\mathcal V_2$ we need to extend the vectors defined on the subspaces $\ket{e_0}\in\mathcal V_1$ and $\ket{e_1},\ket{e_2}\in\mathcal V_2$ to the larger space; define the following vectors
\begin{align}
\label{3dimbvec}
&\ket{e_0}=1~~;~~
\ket{E_0}=\ket{e_0}\oplus\ket{0}= \left[%
\begin{array}{c}
  1   \\
   0 \\
   0
   \end{array}%
\right]\nonumber\\
&\ket{E_1}=\ket{0}\oplus\ket{e_1}= \left[%
\begin{array}{c}
  0   \\
   1 \\
   0
   \end{array}%
\right] ~~;~~
\ket{E_2}=\ket{0}\oplus\ket{e_2}= \left[%
\begin{array}{c}
  0   \\
   1/2 \\
   1/2
   \end{array}%
\right]~
\end{align} 

The dual vectors are given by the transpose, except for $\bra{E^D_1}$ given by
\begin{eqnarray}
\label{eiggenequaldual3}
\bra{E^D_1}= \left [%
\begin{array}{c}
  0 ~,~
   0 ~,~1
   \end{array}\right]
\end{eqnarray}

Let the position operator in the block diagonal space be denoted by $\mathcal X$; from Eq. \ref{xmatrixelements}, since all the elements are dimensionless in the Jordan block representation
\begin{align}
\label{xmatrixelementsdis12}
&\bra {E^D_0} \mathcal X\ket{E_0}=0=\bra {E^D_1}\mathcal X\ket{E_1}=\bra {E^D_2}\mathcal X\ket{E_2}=\bra {E^D_1}\mathcal X\ket{E_2} \nonumber\\
&\bra {E^D_0} \mathcal X\ket{E_1}=1=\bra {E^D_0}\mathcal X\ket{E_2}
\end{align}
and yields the following representation for the Hermitian matrix $\mathcal X$ 
\begin{align}
\mathcal X=\left[%
\begin{array}{ccc}
0&1&1\\
 1& 0 &0  \\
  1& 0& 0 
   \end{array}%
\right]
\end{align}

Since $ \mathcal X$ is dimensionless, its mapping to the coordinate position operator $X$ needs a dimensional scale; let $ \mathcal X=\zeta X$. From Eqs. \ref{ezero}, \ref{eiggenequaldual}, \ref{xmatrixelements} and \ref{xmatrixelementsdis12}
\begin{align}
\label{xmatrixelementsdis}
&1=\bra {E^D_0} \mathcal X\ket{E_1}=\zeta\hat{N}_{00}\sqrt C\bra {\hat{\psi}^D_{00}}X\ket{\psi_{1}}=\zeta\frac{\hat{N}_{00}}{2\omega \sqrt C}\\
&\Rightarrow X=\frac{\hat{N}_{00}}{2\omega \sqrt C}~ \mathcal X~~;~~\zeta=\frac{2\omega \sqrt C}{\hat{N}_{00}}
\end{align}
Extending the Hamiltonian to the $\mathcal V_1\oplus\mathcal V_2$ space yields, from Eq. \ref{jordan22}
\begin{align}
\label{ham3space}
\mathcal H= \mathcal H_1 \oplus \mathcal H_2=\omega \oplus 2\omega\left[%
\begin{array}{cc}
  1 &-1  \\
   0& 1 
   \end{array}%
\right]=\omega\left[%
\begin{array}{ccc}
1&0&0\\
 0& 2 &-2  \\
  0& 0& 2 
   \end{array}%
\right] \nonumber
\end{align}
The evolution kernel is given by
\begin{align}
\exp\{-\tau \mathcal H\}=e^{-2\omega\tau}\left[%
\begin{array}{ccc}
e^{\omega\tau}&0&0\\
 0& 1 &2 \omega\tau \\
  0& 0& 1 
   \end{array}%
\right]
\end{align}

The completeness equation from Eq. \ref{2dimcomple},  has the following extension to $\mathcal V_1\oplus\mathcal V_2$
\begin{align}
\label{3dimcomple}
\mathbb{I}_{3\times 3}&=\ket{E_0}\bra{E_0^D}+ 2\Big[-\ket{E_1}\bra{E^D_1}+\ket{E_2}\bra{E^D_1}+\ket{E_1}\bra{E^D_2}\Big]
\end{align}

In the block-diagonal basis, the propagator is given by
\begin{align}
\label{propghatblockdia}
&\hat{G}(\tau)=\hat{N}^2_{00} \bra{\hat{\psi}^D_{00}}X
e^{-\tau (H-\omega)}X\ket{\hat{\psi}_{00}} \nonumber\\
&=\left(\frac{\hat{N}_{00}}{2\omega \sqrt C}\right)^2\bra{E_0^D} \mathcal X
e^{-\tau (\mathcal H-\omega)}\mathcal X\ket{E_0}
\end{align}
Using the completeness equation given in Eq. \ref{3dimcomple} yields
\begin{align}
\label{propghatdis}
&\hat{G}(\tau)=\frac{\hat{N}^2_{00}}{4\omega^2  C} \nonumber\\
&\times \bra{E_0^D} \mathcal X
e^{-\tau (\mathcal H-\omega)}\left(\ket{E_0}\bra{E_0^D}+ 2\Big[-\ket{E_1}\bra{E^D_1}+\ket{E_2}\bra{E^D_1}+\ket{E_1}\bra{E^D_2}\Big]\right)\mathcal X\ket{E_0} \nonumber\\
&=\frac{\hat{N}^2_{00}}{2\omega^2 C} \bra{E_0^D} \mathcal X
 \Big[-e^{-\omega\tau}\ket{E_1}\bra{E^D_1}+e^{\omega\tau}\ket{E_2(\tau)}\bra{E^D_1}+e^{-\omega\tau}\ket{E_1}\bra{E^D_2}\Big]\mathcal X\ket{E_0} \nonumber\\
 &=\frac{\hat{N}^2_{00}}{2\omega^2 C} e^{\omega\tau} \bra{E_0^D} \mathcal X  \ket{E_2(\tau)}\bra{E^D_1}\mathcal X\ket{E_0}
\end{align}
since
\begin{align*}
%\label{propghat}
&\bra{E^D_1}\mathcal X\ket{E_0} =1=\bra{E_2^D} \mathcal X \ket{E_0} 
\end{align*}
From Eq. \ref{sche2}, the time dependence of $\ket{E_2(\tau)}$ is given by
\begin{align*}
\ket{E_2(\tau)}= e^{-\tau \mathcal H}\ket{E_2} =e^{-2\omega\tau }\left[%
\begin{array}{c}
0\\
  1/2 +\omega \tau  \\
   1/2 
   \end{array}%
\right]~~\Rightarrow~~\bra{E_0^D} \mathcal X  \ket{E_2(\tau)}=1+\omega \tau~~
\end{align*}
and yields, from Eq. \ref{propghatdis}, the expected result for the propagator, namely
\begin{align*}
&\hat{G}(\tau)=\frac{\hat{N}^2_{00}}{2\omega^2 C} e^{-\omega\tau} (1+\omega \tau)=\frac{1}{4\gamma\omega^3} e^{-\omega\tau} (1+\omega \tau)
\end{align*}

A direct derivation can be given using the matrix representation of the evolution kernel; from Eq. \ref{propghatblockdia}
\begin{align*}
%\label{propghat}
&\hat{G}(\tau)=\left(\frac{\hat{N}_{00}}{2\omega \sqrt C}\right)^2\bra{E_0^D} \mathcal X
e^{-\tau (\mathcal H-\omega)}\mathcal X\ket{E_0}
\end{align*}
Using
\begin{align*}
%\label{propghat}
&\mathcal X\ket{E_0}= \left[%
\begin{array}{c}
0\\
  1  \\
   1 
   \end{array}%
\right]  ~~;~~\bra{E_0^D} \mathcal X= \left [%
\begin{array}{c}
  0 ~,~
   1 ~,~1
   \end{array}\right]
\end{align*}
yields, from Eq. \ref{ham3space}, the expected answer
\begin{align*}
%\label{propghat}
&\hat{G}(\tau)=\left(\frac{\hat{N}_{00}}{2\omega \sqrt C}\right)^2e^{-\omega\tau}(2+2\omega\tau)=\frac{1}{4\gamma\omega^3} e^{-\omega\tau} (1+\omega \tau)
\end{align*}

All the $N\times N$ blocks for the Hamiltonian can be analyzed one by one and it can be shown that they are all equal to a corresponding Jordan block matrix. However, the higher order blocks may not be as simple as $\mathcal J_2$ as they can include the direct sum of lower order Jordan blocks.

\section{Conclusions}
The non-Hermitian Euclidean Hamiltonian has all the features of the Minkowski case but with some significant differences. One of the main advantages of the Euclidean formulation is that the both the path integral and the Hamiltonian are well defined, with the Euclidean state space being positive definite.

The equal frequency limit leads to the Euclidean Hamiltonian being equal to a direct sum of Jordan blocks and is a useful example for the study of an irreducibly non-Hermitian system.

\section{Acknowledgment}
I thank Cao Yang for useful discussions and Wang Qinghai for discussing and sharing many of his valuable insights.

%\bibliographystyle{cambridgeauthordate}
%\begin{thebibliography}{99}
\bibliography{../../master_references_all}\label{refs}
\bibliographystyle{plain}
%\label{refs}
%\bibitem{Baaquie}
%\newblock {\it} {\bk}
%\end{thebibliography}
\end{document}